\documentclass[11pt]{article}
\usepackage[english]{babel}
\usepackage{amsmath,amssymb,amsbsy,amstext,amsthm,booktabs,simplewick,exscale,relsize,graphicx,amsfonts,upgreek}
\usepackage{multirow,dcolumn,bm,enumerate,url,hyperref, cite}
\title{\bf Low Scale Inflation at High Energy Colliders and Meson Factories}

\author{Joseph Bramante$^1$, Jessica Cook$^{2,3}$, Antonio Delgado$^1$, Adam Martin$^1$\\
{\small $^1$University of Notre Dame, 225 Nieuwland Hall, Notre Dame, IN, USA}\\
{\small $^2$Arizona State University, 951 Cady Mall, Tempe, AZ, USA}\\
{\small $^3$SUNY Oswego, 7060 Route 104, Oswego, NY, USA}}

\begin{document}
\maketitle

\begin{abstract}
Inflation occurring at energy densities less than (10$^{14}$ GeV)$^4$ produces tensor perturbations too small to be measured by cosmological surveys. However, we show that it is possible to probe low scale inflation by measuring the mass of the inflaton at low energy experiments. Detection prospects and cosmological constraints are determined for low scale quartic hilltop models of inflation paired with a curvaton field, which imprints the spectrum of scalar perturbations observed in large scale structure and on the cosmic microwave background. With cosmological constraints applied, low scale quartic inflation at energies GeV--PeV, can be mapped to an MeV--TeV mass inflaton resonance, discoverable through a Higgs portal coupling at upcoming collider and meson decay experiments. It is demonstrated that low scale inflatons can have detectably large couplings to Standard Model particles through a Higgs portal, permitting prompt reheating after inflation, without spoiling, through radiative corrections to the inflaton's self-coupling, the necessary flatness of a low scale inflationary potential. A characteristic particle spectrum for a quartic inflaton-curvaton pair is identified: to within an order of magnitude, the mass of the curvaton can be predicted from the mass of the inflaton, and vice-versa. Low scale inflation Higgs portal sensitivity targets are found for experiments like the LHC, SHiP, BEPC, and KEKB.
\end{abstract}

\vspace{10cm}

\tableofcontents

\section{Introduction}
Cosmic inflation describes the initialization of our observable universe with remarkably simple elements \cite{Guth:1980zm,Starobinsky:1980te,Linde:1981mu,Mukhanov:1981xt,Albrecht:1982wi}. A scalar inflaton field rolling down its potential is stalled by Hubble friction, so that a ubiquitous negative pressure drives a rapid~ $e^{20}$ -- $e^{60}$ ($20$ -- $60$ efold) increase in the physical distance between spatial points in the universe. While it rolls down its potential, quantum fluctuations of the inflaton source primordial perturbations, whose amplitude is determined by the energy density during inflation and how fast the inflaton rolls. 

Inflationary scalar and tensor perturbations give rise to correlated variations in the primordial plasma of our expanding universe, which eventually manifest as large scale inhomogeneities. The amplitudes of the scalar and tensor power spectrum of these inhomogenities are given by the dimensionless quantities $A_{\rm s}$ and $A_{\rm t}$ respectively \cite{Ade:2015lrj}. Over the preceding decades, measurements of the cosmic microwave background (CMB) and large scale structure have revealed a scalar power spectrum of amplitude $A_{\rm s}^* = (2.206 \pm 0.076) \times 10^{-9} $, with perturbations slightly diminishing over smaller distances, $n_s^* = 0.968 \pm 0.006$ where $n_{\rm s}^*-1 \equiv d {\rm ~ log}~ A_{\rm s}^*/d {\rm ~log}~k$, Refs.~\cite{Hinshaw:2012aka,Ade:2015lrj}. (Quantities with ``$*$" attached are evaluated at an experimentally-determined pivot scale.  In this paper the Planck collaboration's pivot scale is used, $k_* = \frac{0.05}{{\rm  Mpc}}$.)

On the other hand, the size of primordial tensor perturbations have only been bounded from above, and as this bound tightens, so does the bound on the maximum energy scale at which inflation occurred. This is because the energy scale of slow-roll inflation can be directly inferred from the size of the tensor power spectrum,
$
A_{\rm t}^* \simeq 2 V_*/3 \pi^2 M_{\rm p}^4,
$ 
where $V_*$ is the energy density during inflation and $M_{\rm p} \equiv \sqrt{1/8 \pi G}  $ is the reduced Planck mass. The energy scale during slow-roll inflation is often expressed as a combination of the scalar and tensor power spectra ($r \equiv A_{\rm t} / A_{\rm s}$),
\begin{align}
V_*^{1/4} = \left(\frac{3 \pi^2 M_{\rm p}^4 A_{\rm s}^* r_*}{2}\right)^{1/4} \simeq 1.70 \times 10^{16} ~{\rm GeV} \left( \frac{r_*}{0.07} \right)^{1/4},
\label{eq:vbound}
\end{align}
where this expression has been normalized to the $95\%$ confidence bound on $r_*$ reported in \cite{Array:2015xqh}. 

Remarkably, the observation of primordial tensor perturbations could provide some guidance for theories of quantum gravity. A relation known as the Lyth bound indicates that for $r_* \gtrsim 10^{-1}$, the inflaton traversed a field range greater than $M_{\rm p}$ \cite{Lyth:1996im,Boubekeur:2005zm,Efstathiou:2005tq,Easther:2006qu,Bramante:2014rva,Garcia-Bellido:2014wfa}. A super-Planckian inflaton field range (aka large field inflation) indicates that the underlying theory of inflation must, with some symmetry or fixing of parameters, suppress non-renormalizable operators like ``$\phi^6/\Lambda^2$," that otherwise render the theory non-perturbative for all $\Lambda < M_{\rm p}$.  On the other hand, models of inflation with a sub-Planckian inflaton field range (small field inflation) have the advantage of being describable with a low energy effective field theory.  Another reasonable objection to large field inflation, is that many theories predict axions, either as a solution to the strong CP problem \cite{Peccei:1977hh} or as a facet of extra dimensions \cite{Arvanitaki:2009fg}.  Large field inflation often leads to an overabundance of axion dark matter and observationally-excluded axion isocurvature fluctuations \cite{Dine:1982ah,Preskill:1982cy,Hertzberg:2008wr}.

Setting aside theoretical considerations, the bound on tensor perturbations given in Eq.~\eqref{eq:vbound} has already substantially limited viable models of large field inflation. For example, the simple large field potential, $V(\phi)= m^2 \phi^2$, already lies well outside the $2 \sigma$ bound set by BICEP and Keck \cite{Array:2015xqh}. However some well-known large field models with non-standard gravitational couplings, most notably Starobinsky and non-minimally coupled Higgs inflation \cite{Starobinsky:1980te,Bezrukov:2007ep}, could still be found by future astrophysical searches for tensor perturbations.

But a future measurement of tensor perturbations is not guaranteed. The only firm constraint on the inflationary energy density $V_*$ is that it must exceed the energy density required for big bang nucleosynthesis $V_*^{1/4} \gtrsim 10 ~{\rm MeV}$ \cite{deSalas:2015glj,Kawasaki:1999na,Hannestad:2004px}. Thus inflation could have occurred at energies ranging over $V_*^{1/4} \sim 0.01-10^{16}~{\rm GeV}$, corresponding to $r_* \sim 10^{-74} - 10^{-1}$. Planned experiments such as PIXIE and LiteBIRD may probe down to $r_* \sim 10^{-3}$, or equivalently $V_*^{1/4} \sim 5 \times 10^{15}~{\rm GeV}$ \cite{Matsumura:2013aja,Kogut:2011xw}. But it will be challenging for future cosmological experiments to probe much below this, since the intrinsic B-mode polarisation of the CMB in our universe has size $r_* \sim 10^{-7}$, caused by density non-linearities present at recombination, which provide an irreducible background \cite{Fidler:2014oda,Hu:2001fb,Lewis:2001hp,Knox:2002pe,Seljak:2003pn}.

In summary, axion cosmology and an increasingly tight upper bound on the energy scale of inflation point towards low scale inflation. But low scale inflation cannot be confirmed by astrophysical searches for primordial tensor perturbations. Therefore, it is imperative to find non-astrophysical methods to uncover low scale inflation, including terrestrial searches for scalar resonances.

\subsection*{Finding low scale inflation at low energy experiments}

\begin{figure}[t]
\center
\includegraphics[scale=.6]{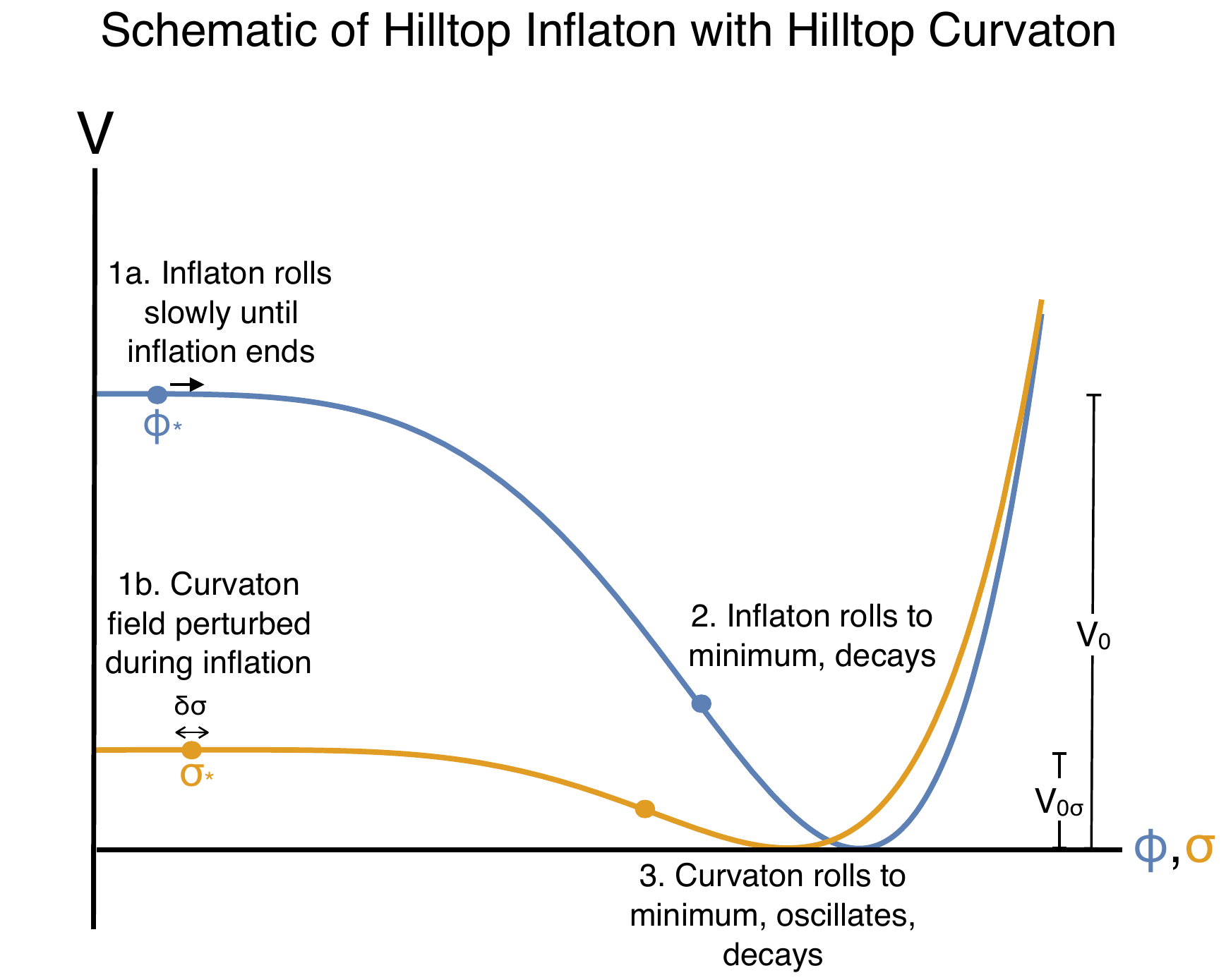}
\caption{Schematic showing inflation driven by a hilltop inflaton, with primordial perturbations provided by a hilltop curvaton. (1a. and 1b.) While the inflaton field $\phi$ slowly rolls to its minimum, the curvaton field is perturbed by de Sitter vacuum fluctuations $\delta \sigma \sim H/2 \pi$, where $H$ is the hubble constant during inflation. (2.) The inflaton field rolls to its minimum and decays. (3.) Sometime later, when the curvaton energy density is the predominant energy density in the universe, the curvaton decays. $V_0$ and $V_{0 \sigma}$ are the change in potential energy of the inflaton and curvaton, respectively. In viable parameter space studied here, $V_{0 \sigma} \ll V_0$.}
\label{fig:schematic}
\end{figure}

To begin unmasking the realm of low scale inflation, this paper shows that particle colliders and meson factories are already poised to probe inflation when $V_*^{1/4} \ll 10^{15}~{\rm GeV}$. This study will focus on a simple case, where the inflaton's dynamics during inflation are determined by a single polynomial term in the Lagrangian (``single-term-dominated"), and find that small field quartic hilltop inflation arising from a $Z_2$-symmetric scalar potential can be discovered through a Higgs portal coupling at upcoming experiments like SuperKEKB, SHiP, and the LHC.  Specifically, the following single-term-dominated hilltop potential is considered,
\begin{align}
V = V_0 - \frac{\lambda_\phi}{4} \phi^4 + \frac{\phi^6}{ \Lambda^2} +V(\sigma) ~ ,
\label{eq:phi4intro}
\end{align}
where $\phi$ is the inflaton field, $V_0$ is a constant energy density, and $V(\sigma)$ is the potential of any other scalar fields, subdominant during inflation, that we address shortly.  Hilltop inflation begins with $\phi$ having a small field value $\phi_*$, then rolling to its minimum at a larger field value, $\phi_{min}$, thereby diminishing the vacuum energy of the universe, $i.e.$ canceling $V_0$. For the hilltop potential in Eq.~\eqref{eq:phi4intro}, a small, negative quartic self-coupling results in a very flat potential around $\phi \sim 0$, permitting slow-roll inflation.  The $ \frac{\phi^6}{ \Lambda^2}$ term is a non-renormalizable effective operator, which stabilizes the potential at its minimum, so that $V(\phi_{min}) \simeq 0$.  Broadly speaking, hilltop inflation captures the dynamics of many models, in which the inflaton rolls to a large field value \cite{Boubekeur:2005zm}.

While a small initial field value ($\phi_*$) and a small self-coupling ($\lambda_\phi \sim 10^{-13}$) permit slow-roll inflation, making either $\phi_*$ or $\lambda_\phi$ too small can result in inflaton perturbations that are too large. On the other hand, making $\phi_*$ or $\lambda_\phi$ too large results in too short an epoch of inflation. These competing considerations, along with methodical computations of the power spectrum, reveal that single-term-dominated small field hilltop potentials cannot both inflate the universe and produce the perturbations observed on the CMB. Therefore, for single-term hilltop inflation, a second ``curvaton" field with potential $V(\sigma)$, can produce the observed CMB perturbations \cite{Lyth:2001nq,Moroi:2001ct,Enqvist:2001zp,Lyth:2002my}.  

A curvaton is a second scalar field displaced from the minimum of its potential during inflation, which rolls to its minimum and decays after the inflaton.  Perturbations to the curvaton's field value during $\phi$-driven inflation become the predominant primordial perturbations in the universe, so long as the curvaton's energy density is the predominant energy density in the universe when it decays.  One simple possibility explored in this study, is that the curvaton has a quartic hilltop potential with the same form as the inflaton, but with a smaller quartic self-coupling. A schematic diagram of quartic hilltop inflation with a quartic hilltop curvaton is given in Figure \ref{fig:schematic}.

Some of this study's findings can be summarized:
\begin{itemize}

\item For the inflaton potential in Eq.~\eqref{eq:phi4intro}, simply mandating $\sim 20-40$ efolds of inflation, sufficiently small scalar primordial perturbations ($P_{\rm \zeta \phi} \lesssim 2.2 \times 10^{-9}$),\footnote{For a curvaton cosmology, the inflaton must produce primordial perturbations smaller than those observed \cite{Langlois:2004nn,Sloth:2005yx}.} and a sub-Planckian cutoff $\Lambda < M_{\rm p}$, creates a predictive map between the energy scale of inflation and the mass of the inflaton at its minimum. For example, $V_*^{1/4} \sim {\rm TeV}$ scale inflation corresponds to an inflaton scalar resonance $m_\phi \sim 30 {\rm ~MeV} - 1~ {\rm GeV}$.

\item Low scale inflation can be detectably coupled to Standard Model particles through a Higgs portal operator ($\lambda_{\phi h} \phi^2 |\Phi|^2$), without upsetting the flatness of the inflaton's potential. A low scale inflaton's self-couplings must be tiny to provide a potential flat enough for inflation, $\lambda_\phi \lesssim 10^{-13}$, which means that any Higgs portal coupling must be small, $\lambda_{\phi h} \lesssim 10^{-6}$, or else spoil the inflaton's self-coupling through radiative corrections.  However, because the vacuum expectation value (VEV) of a quartic hilltop inflaton at its minimum is $10^3 - 10^9~{\rm GeV}$, and Higgs-inflaton mixing scales with the inflaton VEV, ${\rm sin}~\theta_\phi \propto \lambda_{\phi h} v_\phi$, this permits ${\rm sin}~\theta_\phi \sim 0.1$.

\item Using a quartic curvaton potential, and requiring that the curvaton generate the observed scalar perturbation spectrum, fixes the curvaton quartic self-coupling to $1.9 \times 10^{-14} \leq \lambda_{\sigma} \leq 6.9 \times 10^{-14}$ (for the $1 \sigma$ measured values of $n_s^*$ and $A_s^*$ in \cite{Ade:2015lrj}). Using this, and with the inflaton mass specified, the lighter curvaton mass can be predicted (and vice-versa). Similarly, the decay width of the inflaton sets an upper bound on the decay width of the curvaton, and the curvaton decay width sets a lower bound on the inflaton decay width. Therefore, searches for scalars across a range of masses at experiments like the LHC, SHiP, SuperKEKB, BEPC II, and Babar, could identify an inflaton-curvaton pair.

\end{itemize}

Note that it has been appreciated in many contexts that small field inflation requires an extremely flat potential, and as a consequence is naively fine-tuned ($e.g.$ Refs.~\cite{Kumekawa:1994gx,Allahverdi:2006we,Baumann:2007np,Lyth:2007qh,Dine:2011ws}). For Eq.~\eqref{eq:phi4intro}, this manifests as the requirement that the inflaton's quadratic term is negligible during inflation. This study does not seek amelioration of small field fine-tuning, with additional symmetries or a UV theory. 
However, note that the requirement $m^2 \phi_*^2 \ll V_*$ in small field inflation might be compared to the requirement $\phi_*^6 / \Lambda^2 \ll V_*$ in models of large field inflation, for which $\phi_* \gtrsim M_{\rm p}$.

Altogether, this paper demonstrates that low scale inflation can be probed by low energy experiments. Some prior studies have developed similar links between high scale inflation and low energy experiments, in the context of either the Higgs boson or another scalar non-minimally coupled to gravity \cite{Bezrukov:2009yw,Bezrukov:2013fca}, as well as Ref.~\cite{Allahverdi:2010zp}, determining LHC bounds on supersymmetric low-scale inflation. 

The remainder of this paper proceeds as follows. In Section \ref{sec:simplequarticmodel} a simplified low scale quartic model of inflation is introduced, and it is shown that once cosmological constraints are applied, there exists a map between the scale of inflation and the inflaton's mass at its minimum. Section \ref{sec:reheatingconstraints} further constrains the inflaton potential, such that the cosmological epochs of inflation, reheating, radiation, and matter dominated expansion match observations. (Results in Sections \ref{sec:simplequarticmodel}--\ref{sec:reheatingconstraints} apply with or without a curvaton model.) Section \ref{sec:curvaton} studies a quartic curvaton that produces the observed primordial perturbations and identifies viable reheating epochs for a low scale quartic inflaton-curvaton pair, in terms of the average equation of state during reheating ($w_{re}$) and temperature at the end of reheating ($T_{re}$). Section \ref{sec:maps} demonstrates how prior sections can be used to determine an inflaton-curvaton particle spectrum. General prospects for finding low scale inflation through a Higgs portal at colliders and meson factories, and in particular signatures of an inflaton-curvaton spectrum, are explored in Section \ref{sec:inflatonhiggsportal}. In Section \ref{sec:conclusions}, conclusions are presented. Appendix \ref{app:othermodels} discusses the fundamentals, feasibility, and fine-tuning of a variety of small field models, especially small field quartic inflation. Appendix \ref{app:higgsportal} details the Higgs portal paramaterization used in this paper.

\section{Low scale quartic hilltop inflation}
\label{sec:simplequarticmodel}

Inflation occurs when in some region of spacetime, $\ddot{a} >0 $, where $a$ is the scale factor of the universe\footnote{Formally, $a$ is the relative scale of space-like hypersurfaces, $c.f.$ the FLRW metric, $ds^2 = dt^2 - a^2(t) dx^2$.} and $\dot{~} \equiv d/dt$. ``Slow-roll" inflation occurs when, uniformly within a Hubble horizon, defined as $H \equiv \dot{a}/a$, a scalar field is slowly rolling down its potential $V(\phi)$, such that the slow roll parameters $\epsilon$ and $ \eta$ each are much less than unity, that is $\epsilon \equiv -\frac{\dot{H}}{H^2} \simeq \frac{M_{\rm p}^2}{2} \left(\frac{V_{\phi}}{V} \right)^2 \ll 1,$ $|\eta| \equiv M_{\rm p}^2 \left|\frac{V_{\phi \phi}}{V} \right| \leq 1$, $V_{\phi} = \frac {{\rm d} V }{ {\rm d} \phi}$, and $V_{\phi \phi} = \frac{{\rm d}^2 V }{ {\rm d} \phi^2}$. For an introduction to inflation, see $e.g.$~\cite{Baumann:2009ds,Baumann:2014nda}.

This study considers the small field quartic hilltop potential,
\begin{align}
V = V_0 - \frac{\lambda_\phi}{4} \phi^4 + \frac{\phi^6}{\Lambda^2} + V(\sigma),
\label{eq:simplequartic}
\end{align}
where the effective operator $\frac{\phi^6}{\Lambda^2}$ in this potential is negligible during inflation, but is responsible for stabilizing the potential at large field values. The enforcement of the requirement that vacuum energy shut off at the minimum of the potential, for an effective operator with a sub-Planckian cutoff $\Lambda < 10^{19}$ GeV, will provide an important constraint on the inflationary parameter space.  From the standpoint of effective field theory, a term like $\frac{\phi^6}{\Lambda^2}$ is expected if $\phi$ couples to new states with masses $\sim \Lambda$.  For the sake of brevity, this study focuses on potentials without $\phi^3$, and $\phi^5$ terms, which are forbidden if the inflaton potential respects a $Z_2$ symmetry. (Appendix \ref{app:othermodels} addresses such models.)  Throughout, this paper assumes canonical kinetic terms for all fields. 

For hilltop potentials like Eq.~\eqref{eq:simplequartic}, inflation begins when within a Hubble size patch (of radius $\sim 1/H$), $\phi$ uniformly has a small field value, that is close to zero. This is a circumstance which might occur subsequent to a phase transition.\footnote{Note that, in the case of a thermal phase transition, large thermal fluctuations in $\phi$ might prevent $\phi$ from being confined to $\phi \sim 0$. These fluctuations could be avoided by assuming another field coupled to $\phi$, tunnels to a new vacuum, thereby initiating the phase transition. Such a phase transition would then deform the inflaton potential from, for example a quadratic potential to a double-well potential.} With a uniform field value set, the inflaton slowly rolls down its potential until $\epsilon \sim 1$, at which point inflation ends. See Figure \ref{fig:schematic} for a schematic illustration.

As discussed in the introduction, there are simultaneous requirements that the inflaton's potential be flat enough for inflation, but not so flat that it over-produce primordial perturbations.  A quantification of these requirements follows in this section.  (Appendix \ref{app:othermodels} provides further discussion.)  Using these quantifications, it can be shown with a numerical survey of polynomial hilltop models, that single-term-dominated small field hilltop inflation cannot both produce the observed spectrum of scalar primordial perturbations and enough inflation.  Therefore, Section~\ref{sec:curvaton} details how a curvaton field, with potential $V(\sigma)$, would produce the observed perturbations. 

There is some fine-tuning associated with the small field $\phi^4$ hilltop inflation potential given in Eq.~\eqref{eq:simplequartic}. (Fine-tuning is apparently generic for small field models of inflation, Refs.~\cite{Allahverdi:2006we,Baumann:2007np,Lyth:2007qh,Dine:2011ws}.) In the small field quartic hilltop model, tuning arises because at the outset of inflation, when $\phi$ is close to zero, $\phi$'s mass term must be small enough not to upset the flatness of the inflationary potential ($ m_{\phi,*}^2 \ll \lambda_\phi \phi_*^2 $).\footnote{In this paper, $m_\phi$ denotes the mass of the inflaton at its minimum, as determined by its quartic term and vacuum expectation value. The inflaton's quadratic term at $\phi = \phi_* \sim 0$, as determined by bare and loop contributions to its mass, is denoted as $m_{\phi,*}$.} In the absence of some symmetry that forbids the mass term while allowing the quartic, this implies a tuning dependent on the cutoff of the theory, since the quartic term is expected to generate a mass term at one-loop order. One might suppose that an alternative hilltop inflation model which uses a quadratic term, $-m^2 \phi_*^2$, as the dominant term during inflation, could be constructed to be technically natural. Appendix \ref{app:othermodels} surveys hilltop models and shows that, terms higher order in $\phi$, necessary to stabilize such a quadratic potential, re-introduce comparable fine-tuning, assuming an effective field theory with a sub-Planckian cutoff.

For the inflaton potential specified in Eq.~\eqref{eq:simplequartic}, constraints from cosmology shape the allowed parameter space. Hereafter, the requirements that inflation last for $N_* \sim 20-50$ efolds, that vacuum energy vanishes at the minimum of the potential $V(\phi_{min} \sim 0$), and that the inflaton not produce perturbations larger than those on the CMB will be used to pinpoint the mass and vacuum expectation value of the inflaton at its minimum, for a given set of $V_0$ and $\Lambda$.

In the slow-roll limit, for a given $V_0$, $\Lambda$, and initial inflaton field value $\phi_*$, the number of e-folds generated by the quartic hilltop potential, when $\phi $ rolls from $\phi_*$ to the end of slow-roll inflation ($\epsilon \sim 1$) at $\phi_{end}$,
\begin{align}
N_{\rm *} \equiv M_{\rm p}^{-2} \int_{\phi_{end}}^{\phi_*} \frac{V}{V_{\phi}} d \phi \simeq \frac{V_0 }{2 M_{\rm p}^2 \lambda \phi_*^2},
\label{eq:Nefdef}
\end{align}
where $\phi_{end} \gg \phi_*$.

\begin{figure}
\center
\includegraphics[scale=.75]{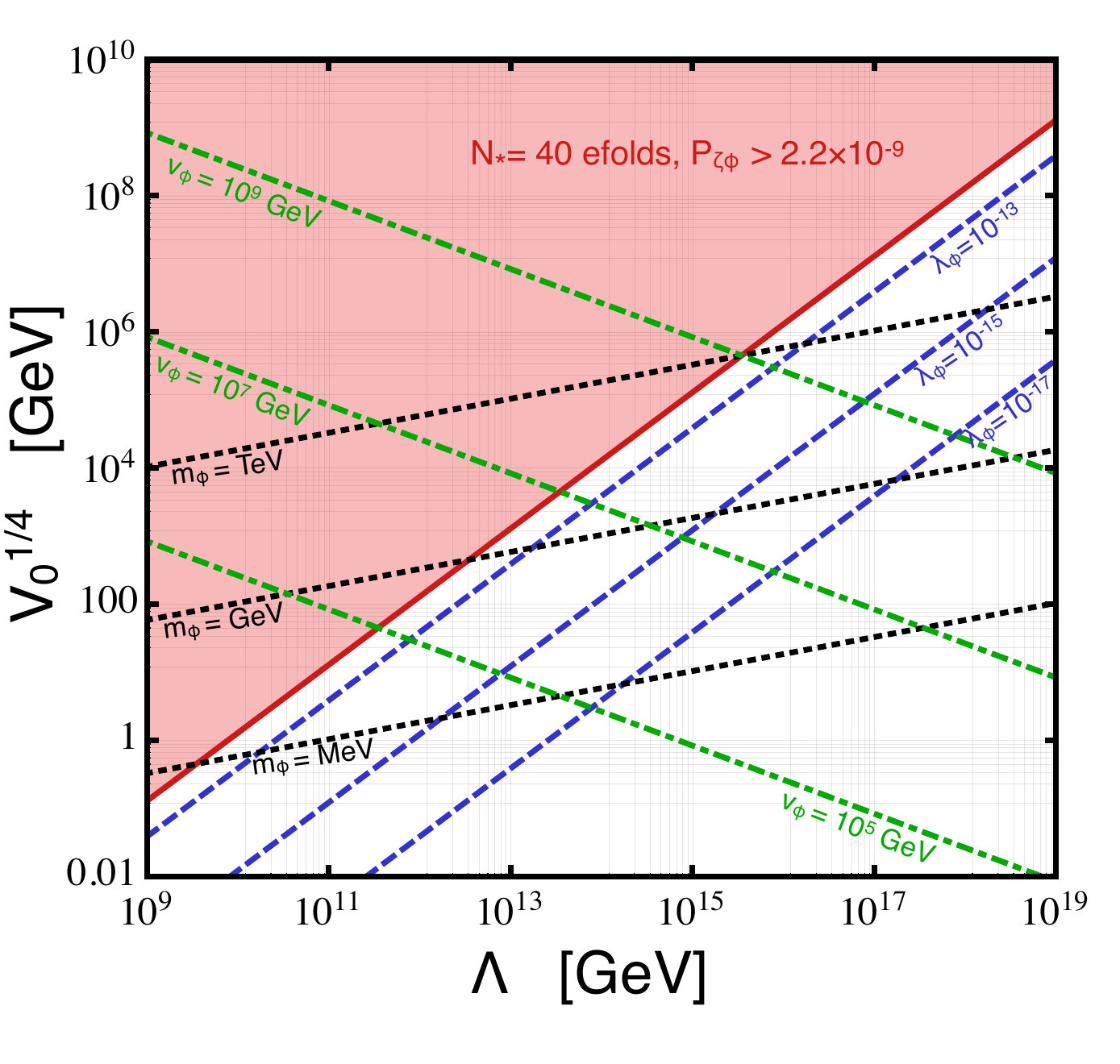}
\caption{Parameter space for the simplified quartic hilltop model given in Eq.~\eqref{eq:simplequartic}. The requirement that the universe inflate by 40 efolds and not produce scalar primordial perturbations larger than those observed on the CMB ($P_{\rm \zeta \phi} \lesssim 2.2 \times 10^{-9}$), excludes the region shaded red. The inflaton's quartic self coupling ($\lambda_\phi$) is fixed in terms of $\Lambda$ and $V_0$ by requiring that the inflaton's potential, Eq.~\eqref{eq:simplequartic}, is zero at its minimum, yielding the relation given in Eq.~\eqref{eq:simplam}. The inflaton's mass ($m_{\phi}$) and vacuum expectation value ($v_{\phi}$) at its minimum are indicated by dotted and dotted-dashed lines.}
\label{fig:V0Lambda}
\end{figure}

In order that $V(\phi)$ does not contribute to the cosmological constant or create anti-de Sitter collapse after $\phi$ rolls to its minimum, it is required of Eq.~\eqref{eq:simplequartic} that $V(\phi_{\rm min}) = 0,$ where $\phi_{\rm min}$ is the value of $\phi$ at the minimum of the potential. (Given the vacuum energy observed in our universe, technically this requirement could be relaxed to $V(\phi_{\rm min}) \lesssim {\rm meV^4}$, but this would not change the inflaton's couplings enough to alter results.) This fixes $\lambda_{\phi}$ in terms of $V_0$ and $\Lambda$. Specifically, $V(\phi_{\rm min}) = 0$ implies 
 \begin{align}
\lambda_{\phi} = 6 \left( \frac{2 \, V_0}{\Lambda^4} \right)^{\frac{1}{3}}.
\label{eq:simplam}
\end{align}

With this expression for $\lambda_{\phi}$, the number of efolds can be expressed in terms of $V_0$ and $\Lambda$,
\begin{align}
N_{\rm *} \simeq \frac{V_0^{\frac{2}{3}} \Lambda^{\frac{4}{3}} }{12 \cdot 2^{\frac{1}{3}} M_{\rm p}^2 \phi_*^2}.
\label{eq:Nefforsimpmodel}
\end{align}
Similarly, using Eqs.~\eqref{eq:Nefdef} and \eqref{eq:simplam}, the vacuum expectation value of the inflaton at its minimum,
\begin{align}
v_\phi = \left(2 V_0 \Lambda^2 \right)^{1/6} \simeq 5~{\rm PeV} \left(\frac{V_0}{({\rm TeV})^4} \right)^{1/6}\left( \frac{\Lambda}{10^{14}~{\rm GeV}} \right)^{1/3},
\label{eq:vphi}
\end{align}
along with the mass of the inflaton at its minimum,
\begin{align}
m_{\phi} = \sqrt{12}\left(2 \frac{V_0}{\Lambda} \right)^{1/3}\simeq 2 ~{\rm GeV} \left(\frac{V_0}{({\rm TeV})^4} \right)^{1/3}\left( \frac{10^{14}~{\rm GeV}}{\Lambda} \right)^{1/3},
\label{eq:mphi}
\end{align}
can also be determined as a function of $\Lambda$ and $V_0$.
 
It should be required that the spectrum of scalar perturbations produced by the inflaton in Eq.~\eqref{eq:simplequartic} not be larger than that observed on the CMB  ($P_{\rm \zeta \phi} \lesssim 2.2 \times 10^{-9}$). As detailed in Section~\ref{sec:curvaton}, a curvaton is assumed to produce the perturbations observed on the CMB. However, if the inflaton perturbations are too large, these can be transferred via gravitational coupling, increasing the curvaton's perturbations \cite{Langlois:2004nn,Sloth:2005yx}. Using slow-roll formulae for scalar primordial perturbations, $\phi$'s perturbations should be subdominant,
\begin{align}
P_{\zeta  \phi} \simeq \frac{V}{24 \pi^2 M_{\rm p}^4 \epsilon} = \frac{V_0^3}{12 \pi^2 M_{\rm p}^6 \lambda_\phi^2 \phi_*^6} \leq A_s.
\label{eq:Asdef}
\end{align} 
In the limiting case of a Planck-scale cutoff $\Lambda = 1.2 \times 10^{19}~{\rm GeV}$, scalar primordial perturbations are small enough to accommodate observation so long as $\lambda_\phi \lesssim 10^{-13}$ and $V_0 \lesssim 10^{9}$ GeV. This is the maximum energy scale for small field quadratic hilltop inflation, given the observed primordial power spectrum, $A_{\rm s}^* \simeq 2.2 \times 10^{-9}$. To show this, first, $\phi_*$ is fixed by Eq.~\eqref{eq:Nefforsimpmodel} and the requirement that inflation last for $\sim 40$ efolds. Then $\phi_*$ is substituted into Eq.~\eqref{eq:Asdef}, and the relation of Eq.~\eqref{eq:simplam} is incorporated. Altogether, the requirements of sufficient efolds, a small enough primordial power spectrum, and that the inflaton's potential energy vanish at its minimum imply
\begin{align}
\lambda_{\phi} \leq 5.1 \times 10^{-13} \left(\frac{P_{\rm \zeta \phi}}{2.2 \times 10^{-9}} \right) \left(\frac{40}{N_{*}} \right)^3,
\label{eq:Asquarticbound}
\end{align} 
which can be re-cast as a bound on $\Lambda$ and $V_0$ with Eq.~\eqref{eq:simplam},
\begin{align}
V_0^{1/4} \leq 1.3 \times 10^{-10} \; \Lambda \left(\frac{P_{\rm \zeta \phi}}{2.2 \times 10^{-9}} \right)^{3/4} \left(\frac{40}{N_{*}} \right)^{9/4}.
\label{eq:V0Lambound}
\end{align}

In Figure \ref{fig:V0Lambda}, parameter space is shown in terms of $V_0$ and $\Lambda$, consistent with $\sim 40$ e-folds of inflation and sufficiently small inflaton perturbations, $P_{\rm \zeta \phi} \lesssim 2.2 \times 10^{-9}$. This plot additionally demonstrates that, assuming the minimal $Z_2$ symmetric hilltop potential of Eq.~\eqref{eq:simplequartic}, the mass of the inflaton at its minimum predicts the scale of inflation to within an order of magnitude. For example, a GeV mass inflaton implies $V_0^{1/4} \sim 0.3-10$ TeV. This raises the possibility of inferring the scale of low energy inflation by measuring the mass of the inflaton at a low energy experiment, as detailed hereafter.

Note that so far, no assumptions about the curvaton sector has been made, and so the preceding relationship between the sub-Planckian effective operators stabilizing an inflaton, and its mass and vacuum expectation value at its minimum, could be applied to any hilltop inflaton, with a weak-enough self-coupling to generate enough efolds of inflation, but not so weak as to over-produce primordial perturbations.

\section{Cosmological consistency and low scale quartic hilltop inflation}
\label{sec:reheatingconstraints}

This section shows that requiring the shrinkage of the comoving horizon during inflation, match its subsequent expansion during reheating, radiation-dominated, and matter-dominated expansion ($e.g.$ \cite{Cook:2015vqa,Dai:2014jja,Martin:2010kz}), provides another constraint on plausible combinations of $V_0$, $N_*$, and $\Lambda$. The relevant formalism is derived and extended to accommodate a curvaton, then applied to low scale quartic hilltop inflation. The key point is that after restricting the equation of state and temperature of reheating to plausible values ($w_{re} \sim 0$ -- $\frac{1}{3}$ and $T_{re}\sim 4.7 ~{\rm MeV}$ -- $V_0^{1/4}$ respectively), inflation has both a minimum and maximum corresponding duration. These considerations restrict the number of efolds of inflation ($N_*$) to a narrow window of possible values, for a given set of $(V_0,\Lambda)$.  One use of this narrowed range of plausible efold values, is to help tighten maps between inflaton and curvaton parameters in Sections \ref{sec:maps} and \ref{sec:inflatonhiggsportal}. 

\subsection{Cosmological consistency for low scale inflaton-curvaton models}

One advantage of inflationary cosmology is that it explains the uniformity of the observable universe: during an epoch of inflation, the comoving horizon ($\equiv (aH)^{-1}$) of the universe shrinks. As a result, an observer sees a smaller casually-connected volume in the future, in contrast to an observer watching a universe dominated by matter or radiation, which grows to a larger  causally-connected volume in the future. It is well-established that our universe underwent a period of radiation and matter dominated expansion, implying that the most distant regions presently observed were once far outside of causal contact. 
Inflation serves to drive pieces of the present causally-connected universe out of causal contact, before radiation and matter dominated expansion, thereby allowing for a present-day homogeneous universe.

This also means a consistent inflationary cosmology requires that the shrinkage of the comoving horizon during inflation, is equal to the growth of the comoving horizon after inflation. The amount the comoving horizon grows after inflation will depend upon the equation of state of the expanding universe -- though eventually the universe must become radiation dominated to accommodate big bang nucleosynthesis, after which the growth of the comoving horizon can be determined from observation. Bounds on inflaton models from a consistent cosmology have been explored in \cite{Martin:2010kz,Martin:2013tda,Dai:2014jja,Munoz:2014eqa,Cook:2015vqa}. See Figure 1 of Ref.~\cite{Cook:2015vqa} for an illustrative schematic.

To bound inflation using a consistent cosmology, we begin by considering modes relevant to observations of the CMB.  During inflation, when the comoving horizon ($1/aH$) shrinks to a size smaller than $1/k$, modes of size $k$ depart the comoving horizon. The mode corresponding to the CMB pivot scale has already been defined as $k_*$, and the Planck collaboration uses $k_* = 0.05 \frac{1}{{\rm Mpc}}$ in their analyses. Therefore, with a pivot scale of $k_*$, Planck's measurements of $n_s^*$ and $A_s^*$ are determined by inflationary dynamics occurring when the comoving scale was of size $\sim 1/k_*$, in other words $k_*=a_*H_*$. 
\\

\begin{figure}[t!]
\center
    \includegraphics[width=.85\textwidth]{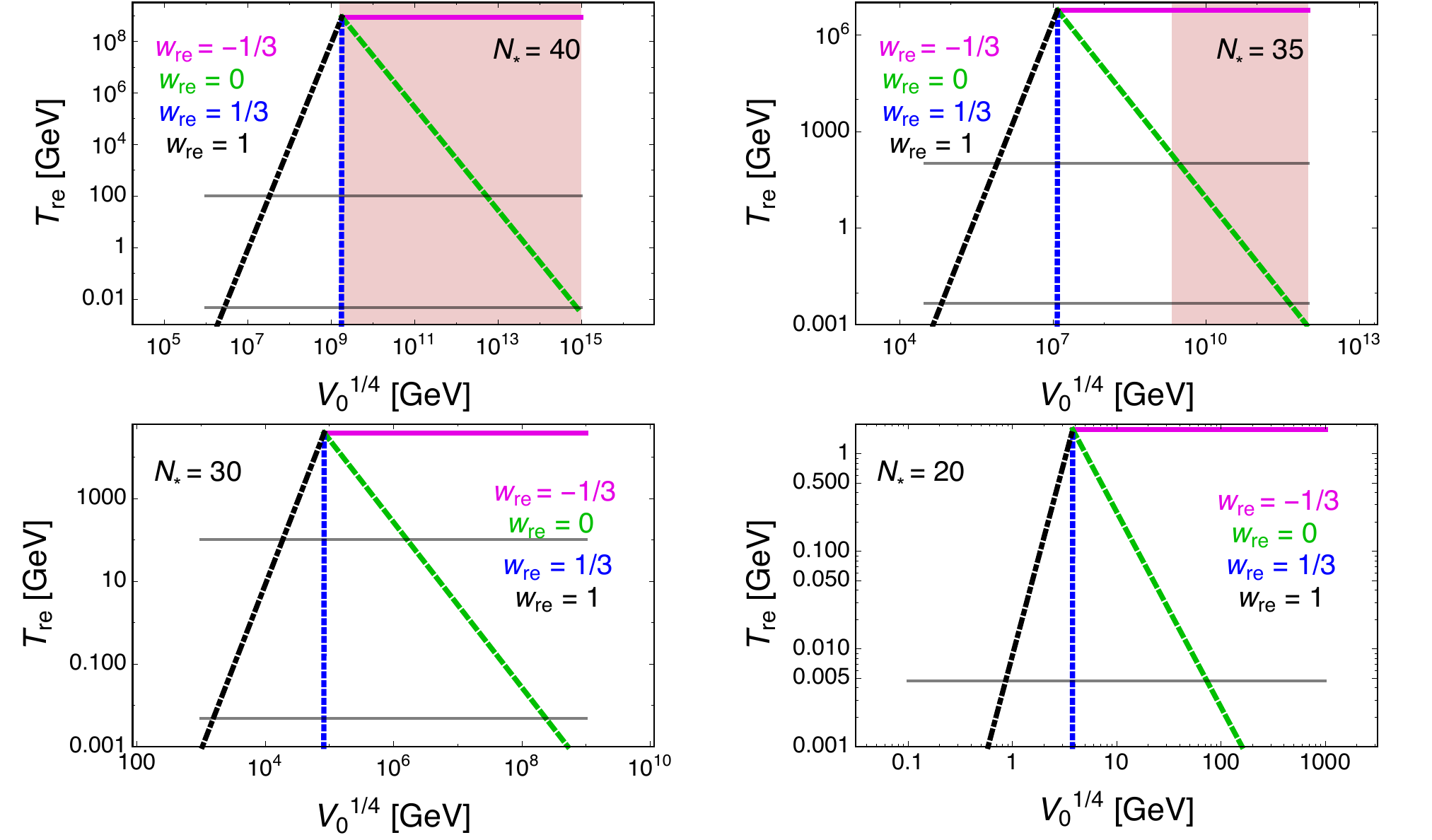}
    \caption{Constraints are given for inflationary energy density ($V_0$) and initial reheating temperature ($T_{re}$) for a cutoff $\Lambda = 10^{19}~{\rm GeV}$, assuming quartic hilltop inflation, defined in Eq.~\eqref{eq:simplequartic}. The solid pink, dashed green, dotted blue, and dotted-dashed black lines indicate a post-inflation average equation of state of $w_{re} = 1,\frac{1}{3},0,-\frac{1}{3}$, respectively, as described in the text. As in Figure \ref{fig:V0Lambda}, the region shaded red is excluded because the inflaton produces primordial perturbations that are too large. The bottom horizontal line marks a BBN reheat temperature; space below this line is excluded. The upper horizontal marks the approximate electroweak symmetry breaking temperature (100 GeV). As $N_*$ decreases in each panel, so does the maximum allowed reheating temperature and $V_0^{1/4}$ values, contained within a wedge of sensible equation of state values, $w_{re} = [0,\frac{1}{3}]$. Note that this leads to substantially different y-axis ($T_{re}$) and x-axis ($V_0^{1/4}$) ranges, as $N_*$ is varied.}
 \label{fig:reheatcons1}
\end{figure}

\begin{figure}[t!]
\center
    \includegraphics[width=.85\textwidth]{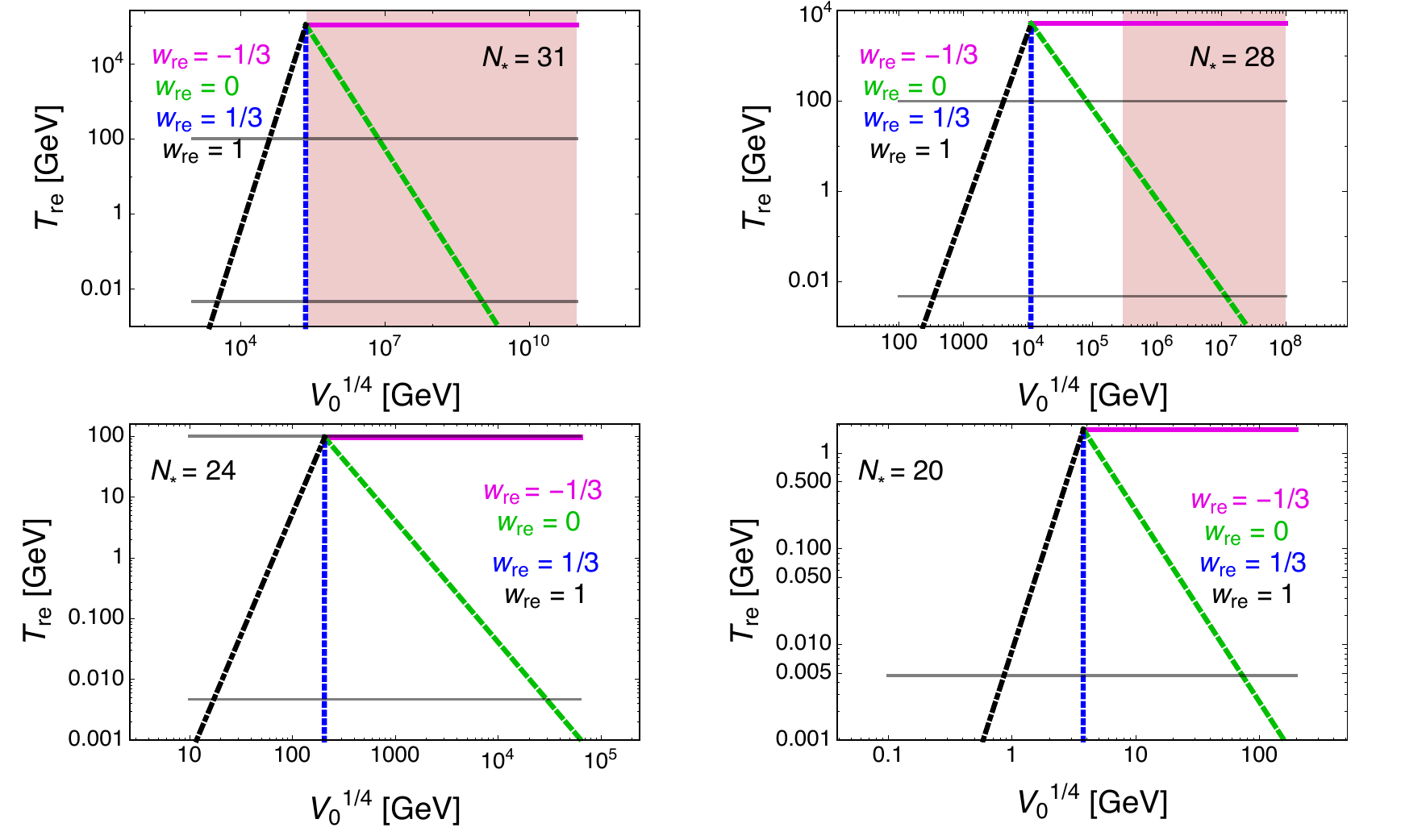}
    \caption{The same as in Figure \ref{fig:reheatcons1} but for $\Lambda = 10^{15}$ GeV.}
\label{fig:reheatcons2}
\end{figure}

\begin{figure}[h]
\center
    \includegraphics[width=.85\textwidth]{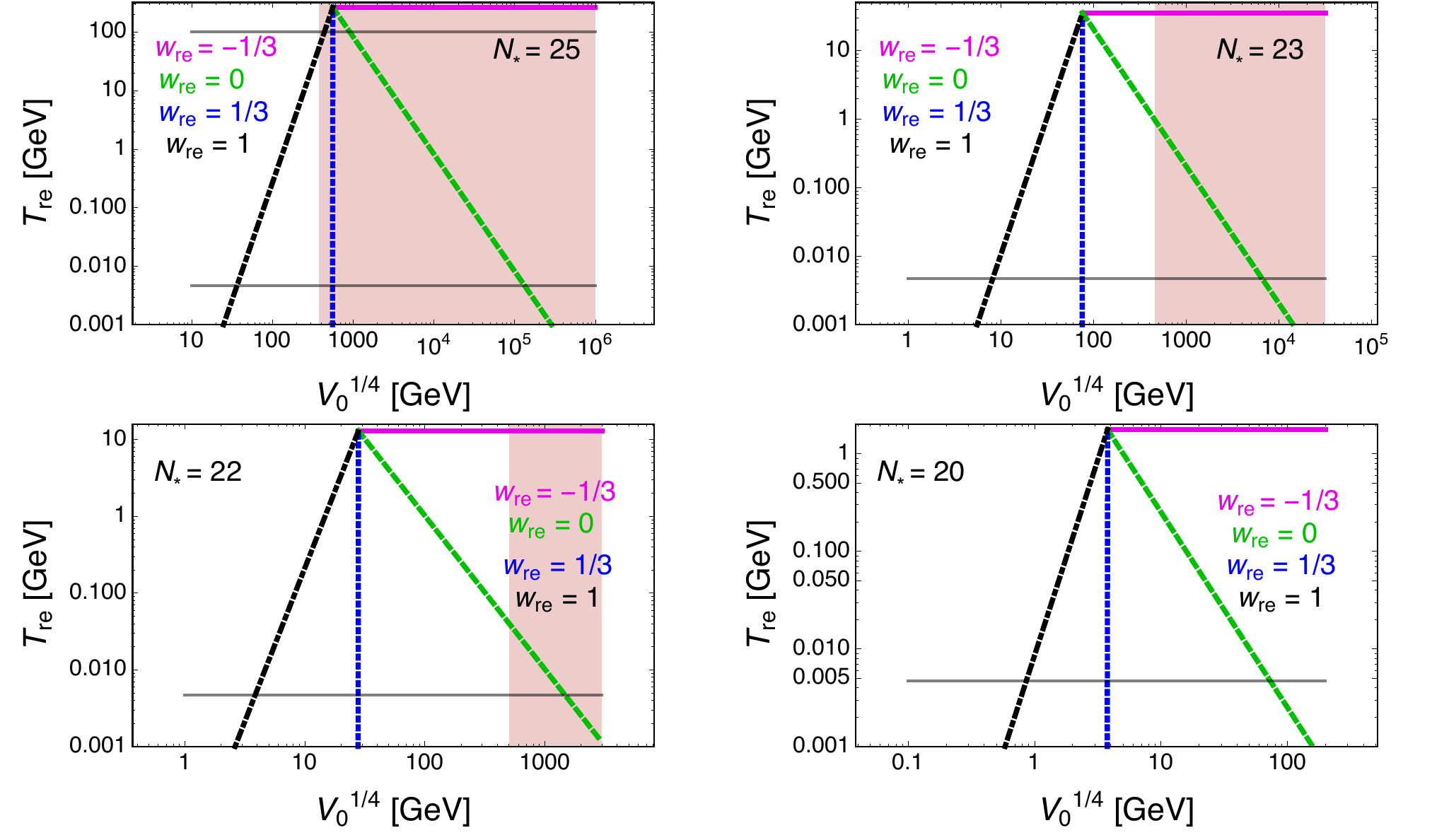}
    \caption{The same as in Figure \ref{fig:reheatcons1} but for $\Lambda = 10^{12}$ GeV.}
\label{fig:reheatcons3}
\end{figure}

Using the relation $k_* = a_* H_*$, multiplied by the present day comoving scale,
\begin{align}
\frac{k_*}{a_0 H_0} = \frac{a_*}{a_{end}} \frac{a_{end}}{a_{re}} \frac{a_{re}}{a_{eq}} \frac{a_{eq} H_*}{a_0 H_0},
\label{eq:expanded}
\end{align}
where each $a$ is the physical scale of the universe at the transition between cosmological epochs.  $a_{end}$ is the size of the universe when inflation ends ($\epsilon \geq 1$). In a number of studies, namely Refs.~\cite{Dai:2014jja,Munoz:2014eqa,Cook:2015vqa}, $a_{re}$ was defined as the scale of the universe after the inflaton has finished decaying, also called the end of reheating. However, our model utilizes a curvaton field, which will decay into radiation sometime after the inflaton decays. In this case, the epoch of reheating lasts until the curvaton decays, and so we define $a_{re}$ as the physical scale of the universe after the curvaton has finished decaying, at which time the universe begins radiation-dominated expansion. As the universe cools, matter and radiation will come to equally populate the energy of the universe when the physical scale is of size $a_{eq}$. Following standard conventions, $a_0$ denotes the present-day scale of the universe.

Equation \eqref{eq:expanded} can be rewritten using the identity $\frac{a_f}{a_i} = e^{\Delta N}$,
\begin{align}
\frac{k_*}{a_0 H_0} = e^{-N_*}  e^{- N_{re}} e^{-N_{RD}} \frac{a_{eq} H_*}{a_0 H_0},
\label{eq:expefolds}
\end{align} 
where $e^{N_{re}} \equiv \frac{a_{end}}{a_{re}}$ is the number of efolds between the end of inflation and when the curvaton finishes decaying, and $e^{N_{RD}} \equiv \frac{a_{re}}{a_{eq}}$ is the number of efolds between the time of curvaton decay and matter-radiation equality. 

Note that in the preceding, the equation of state of the universe $w_{re}$ has not been specified for the period when $a_{end}$ grows to size $a_{re}$. The equation of state during that era of expansion will depend on the decay rate and energy density of both the inflaton and curvaton. More precisely, in a straightforward curvaton cosmology, the inflaton decays more rapidly than the curvaton after inflation, so that while inflaton-sourced radiation energy density dilutes like $a^{-4}$ as the universe expands, the un-decayed curvaton field behaves more nearly like matter, $w \sim 0$, so that its energy density dilutes like $\sim a^{-3}$. Then as the universe expands and $a$ increases, the curvaton's energy density grows to exceed the inflaton's energy density. Sometime after it comes to dominate the energy density of the universe, the curvaton decays. This process results in a universe with primordial perturbations that depend (almost) solely on the curvaton's field perturbations during inflation \cite{Lyth:2001nq}. Hereafter we refer to this entire epoch ($a_{end} \rightarrow a_{re}$) as the era of reheating. As we will see, Eq.~\eqref{eq:expefolds} can be used to relate the parameters of inflation to those of reheating. 

First, the Friedmann equations can be combined to yield an expression relating the initial and final energy densities of an expanding, isotropic universe,  
$
\rho_{re} = \rho_{end} \, e^{- 3 N_{re} \left(1+  w_{re} \right)}.
$ 
For the case we are interested in, here $\rho_{re}$ is the total energy density at the end of reheating, $\rho_{end}$ is the total energy density at the end of inflation, and $w_{re} $ is the average equation of state during reheating, during which time the energy density of the universe flips from being inflaton-dominated to curvaton-dominated. As already mentioned, to avoid largely excluded isocurvature perturbations, the energy density at the end of reheating, $\rho_{re}$, must be predominantly energy density sourced by the decayed curvaton.

Next we re-express the energy density at the end of reheating as a temperature, using the number of relativistic degrees of freedom,
$
\rho_{re} \sim \frac{\pi^2}{30} g_{re} T_{re}^4,
$
see $e.g.$ \cite{Mukhanov:2005sc}. Assuming conservation of entropy after reheating and using the fact that the relativistic species of the present-day universe are photons and neutrinos,
\begin{align}
a_{re}^3 g_{re} T_{re}^3 = a_0^3 \left(2 T^3_{0 } + 6 \cdot \frac{7}{8} T^3_{\nu 0} \right)
\label{eq:aregre}
\end{align}
where $T_0 \simeq 2.725 $K and $T_{\nu 0} = \left(\frac{4}{11} \right)^{1/3} T_0$. Putting Eqs.~\eqref{eq:expefolds}--\eqref{eq:aregre} together, we find that so long as $w_{re} \neq \frac{1}{3}$,\footnote{If $w_{re} = \frac{1}{3}$ instead of Eq.~\eqref{eq:Nre},
\begin{align}
- \frac{1}{4} {\rm log}\left(\frac{90}{2 \pi^2 g_{re}}\right) - \frac{1}{3} {\rm log}\left(\frac{11 g_{re}}{43} \right) - {\rm log}\left(\frac{k_*}{a_0 T_0} \right)  = {\rm log} \left(\frac{V_{end}^{\frac{1}{4}}}{H_*}\right) + N_*,
\end{align}
however in a straightforward curvaton cosmology, $w_{re} < \frac{1}{3}$ during reheating, so that the curvaton's energy density grows to exceed the inflaton's energy density.} the amount the comoving horizon grows (in efolds) and the temperature at which the universe reheats are given by
\begin{align}
N_{re} = \frac{4}{\left(1- 3 w_{re}\right)} \left[- \frac{1}{4} {\rm log}\left(\frac{90}{2 \pi^2 g_{re}}\right) - \frac{1}{3} {\rm log}\left(\frac{11 g_{re}}{43}\right) - {\rm log}\left(\frac{k_*}{a_0 T_0}\right) - {\rm log} \left(\frac{V_{end}^{\frac{1}{4}} } {H_*}\right) - N_* \right]
\label{eq:Nre}
\end{align}
and
\begin{align}
T_{re} = \left(\frac{43}{11 g_{re}} \right)^{\frac{1}{3}} \frac{a_0 T_0}{k_*} H_* e^{-N_*} e^{-N_{re}} \, .
\label{eq:Tre}
\end{align}
In the preceding expressions, the substitution $\rho_{end} = V_{end} \simeq V_0$ can be made, because the inflaton's energy density will be the predominant energy density in the universe at the end of inflation. In computations that follow, $H_*$ is similarly determined by the inflaton's energy density during inflation, $i.e.$ $3H_*^2 \simeq V_0/M_{\rm p}^2 $, and the number of relativistic degrees of freedom in the Standard Model is taken to be $g_{re} \sim 100$. Using the Planck defined pivot scale, and the standard normalization for $a_0$, in Eq.~\eqref{eq:Tre} $k_*/a_0 \rightarrow 0.05({\rm Mpc})^{-1}$.

\subsection{Results for small field quartic hilltop inflation}

In Figures \ref{fig:reheatcons1}--\ref{fig:reheatcons3}, we plot the implied final reheating temperature $T_{re}$ for indicated values of $N_*$, by substituting Eq.~\eqref{eq:Nre} into Eq.~\eqref{eq:Tre}. Parameter space lying between the solid red and dotted-dashed black lines, where these correspond to an average equation of state between $- \frac{1}{3}$ and 1, is technically permitted, but more realistically one should only consider parameter space lying between the dotted blue and dashed green lines, which correspond to a matter- or radiation-like equation of state during reheating. The bottom horizontal line marks the temperature at big bang nucleosynthesis, $T_{BBN} \geq 4.7 ~$MeV -- any realistic cosmology must reheat at a higher temperature \cite{deSalas:2015glj}. The upper horizontal line marks an estimate for the temperature of electroweak symmetry breaking $\sim 100 GeV$. A cosmology which assumes electroweak baryogensis would need to occupy parameter space above this line. 

The red shaded regions are excluded by requiring that inflaton perturbations not be too large, as discussed around Eq.~\eqref{eq:V0Lambound}; the same bound is indicated with red shading in Figure \ref{fig:V0Lambda}. Altogether, Figures \ref{fig:reheatcons1}--\ref{fig:reheatcons3} indicate a number of constraints on viable quartic hilltop parameter space. For $\Lambda = 10^{19},10^{15},10^{12}$ GeV, no consistent cosmology can be constructed when $N_* > 40,31,25$ efolds, respectively. More generally, as the cutoff is lowered from $\sim 10^{19}$ GeV to $10^{12}$ GeV, the number of efolds consistent with a given inflationary energy density $V_0$ also shrinks; this can also be seen directly from Eq.~\eqref{eq:Nefdef}. Further inspecting $\Lambda = 10^{15}$ GeV parameter space, we find that requiring a reheat temperature above the scale of electroweak symmetry breaking, restricts the number of efolds to $N_*=24-31$. 

\section{Quartic hilltop curvaton perturbations and cosmology}
\label{sec:curvaton}
 
Small field quartic hilltop inflation would not generate the observed spectrum of primordial perturbations observed in our universe (see Section \ref{sec:simplequarticmodel} and Appendix \ref{app:othermodels}). Therefore, a low scale quartic hilltop inflaton requires an additional curvaton field ($\sigma$) to produce the observed spectrum of perturbations. In a curvaton cosmology, the inflaton and its decay products dominate the universe's energy density after inflation, but eventually, the curvaton's energy density grows to exceed the inflaton's energy density in the expanding universe. At this time, the curvaton decays, and the perturbations of the curvaton field become the predominant primordial perturbations observed in the universe. In the case of high scale inflation, a simple curvaton potential like $V(\sigma) = m^2 \sigma^2$ can be employed, but such curvaton potentials cannot produce the observed primordial perturbations in the case of low scale inflation. The hilltop curvaton is arguably the simplest practicable curvaton for low scale inflation \cite{Dimopoulos:2004yb}, and so a quartic hilltop inflaton is employed here.

In the remainder of this section, perturbations from a quartic hilltop curvaton are detailed, along with the application of a consistent history for low scale curvaton cosmology. Once all cosmological constraints are applied, a limited range of reheating efolds ($N_{re}$), equations of state ($w_{re}$), and inflationary energy densities ($V_0$) are permitted for a given cutoff ($\Lambda$) -- this relationship is surveyed in Figure \ref{fig:rainbow}.

We begin with a curvaton potential that is identical to the quartic hilltop inflaton potential. 
\begin{align}
V \left(\sigma \right) =V_{0 \sigma} - \frac{\lambda_\sigma}{4} \sigma^4 + \frac{\sigma^6}{\Lambda^2},
\label{eq:curvatonquartic}
\end{align}
where for simplicity we assume the cutoff for the curvaton effective operator ($\Lambda$) is the same as that of the inflaton. Also for the sake of simplicity, we assume that $\phi$ and $\sigma$ only couple substantially through gravity. As will be shown in this section, requiring the curvaton produce the observed spectrum of scalar perturbations, $i.e.$ $n_s^* \sim 0.97$ and $A_s^* \sim 2.2 \times 10^{-9}$, will be enough to uniquely determine $\lambda_\sigma$ and the curvaton's initial field value, $\sigma_*$.  As for the inflaton, $V_{0 \sigma}$ is determined by the curvaton field's self-couplings, and equivalently its field value at is minimum $\sigma_{min} = \sqrt{\frac{\lambda_{\sigma}}{6}} \Lambda$, such that $V(\sigma_{min})=0$.  As illustrated in Figure \ref{fig:schematic}, this section will show that in viable curvaton parameter space, the curvaton will cancel a much smaller portion of vacuum energy as it rolls to its minimum, $V_{0 \sigma} \equiv \lambda_{\sigma}^3 \Lambda^4/432 \ll V_0$.  Hence to good approximation one is justified in neglecting the curvaton's contribution to vacuum energy during inflation.

Hereafter we provide a self-contained derivation of the quartic hilltop curvaton's perturbation spectrum. In the standard curvaton scenario \cite{Lyth:2001nq,Lyth:2002my}, at the onset of inflation, the curvaton is fixed to some field value $\sigma_*$ such that it is slowly rolling, $|V_{\sigma \sigma}| \ll H$, and so the equation of motion for the curvaton perturbations is given by
\begin{align}
\delta \ddot{\sigma} + 3 H \delta \dot{\sigma} + \frac{k^2}{a^2} \delta \sigma = 0 \, ,
\end{align}
which in turn implies that for modes which have exited the comoving horizon ($i.e.$ in the limit $k \ll  aH$ \cite{Baumann:2014nda}), $
\langle \delta \sigma^2 \rangle = \frac{H^2}{2 k^3}.$

We calculate the curvaton power spectrum using $\zeta = - H \frac{\delta \rho}{\dot{\rho}}$, where $\zeta$ parameterizes the scalar perturbations of a scalar field in de Sitter space, and $\rho$ is the energy density of said field. One can define separate $\zeta_i$ for each scalar field $i$ present during inflation. Each $\zeta_i$ will be seperately conserved outside the horizon, provided that the fields only interact gravitationally with each other and have canonical kinetic terms. A violation of either of these conditions would result in time evoluation of $\zeta_i$ on superhorizon scales \cite{Malik:2004tf}. Note that the conservation of each scalar field's perturbations can be applied to other multifield inflationary scenarios. The main difference in the case of a ``curvaton" field, is that the curvaton is not determining how quickly inflation is ending, which alters the spectrum of perturbations it induces on the CMB (relative to an ``inflating" field).

With these provisos, $\zeta_{\sigma} =- H \frac{\delta \rho_{\sigma}}{\dot{\rho}_{\sigma}} $. To unpack this expression, we first expand the curvaton potential using $\sigma = \sigma_0 + \delta \sigma$,
\begin{align}
V(\sigma) = V_{0 \sigma} - \frac{\lambda_{\sigma}}{4} \sigma^4 + \frac{1}{\Lambda^2} \sigma^6 \ = V_{0 \sigma} - \frac{\lambda_{\sigma}}{4} (\sigma_0^4 + 4 \sigma_0^3 \, \delta \sigma + ... ) + ... \, .
\end{align}
Where as with the inflaton, the inflationary field values of $\sigma$ are small enough, that the $\phi^6/\Lambda^2$ term can be dropped so that $\delta \rho_{  \sigma} = - \lambda_{\sigma} \sigma_0^3 \, \delta \sigma$. With $\delta \rho_\sigma$ specified, we can calculate the curvaton's power spectrum,
\begin{align}
P_{\rm \zeta \sigma} \equiv \frac{k^3}{2 \pi^2} \langle \zeta_{\sigma}^2 \rangle \, .
\end{align}
Inserting $\zeta_{\sigma} =- H \frac{\delta \rho_{\sigma}}{\dot{\rho}_{\sigma}} $, $\langle \delta \sigma^2 \rangle = \frac{H^2}{2 k^3}$, and $\delta \rho_{  \sigma} = - \lambda_{\sigma} \sigma^3 \, \delta \sigma$ into this expression yields
\begin{align}
P_{\rm \zeta \sigma}  = \frac{ H^4  \lambda^2_{\sigma} \sigma^6}{4 \pi^2 \dot{\rho}_{\sigma}^2}  \, .
\label{eq:pzetaquart}
\end{align}
The change in time of curvaton energy density, $\dot{\rho}_{\sigma}$, can be calculated using the slow-roll formula: $3 H \dot{\sigma} = -V_\sigma$ implies
$
\dot{\rho}_{\sigma} \approx  -\frac{\lambda_{\sigma}^2 \sigma^6 }{3 H} \, .
$
Inserting this into Eq.~\eqref{eq:pzetaquart} leads to
\begin{align}
P_{\rm \zeta \sigma}  = \frac{ 9 H^6 }{4 \pi^2 \lambda_{\sigma}^2 \sigma^6} \, .
\label{eq:pzetafinquart}
\end{align}

Surveys of the CMB completed by the WMAP and Planck satellites require that $P_{\rm \zeta \sigma} \simeq 2.2 \times 10^{-9}$. These experiments have also measured the scale dependence of primordial scalar perturbations, defined here as $n_s -1 = \frac{d ~{\rm log}~ P_{\zeta}}{d ~{\rm log}~ k}$. Writing $n_s$ as a derivative of the power spectrum with respect to time, again using the relation that a comoving momentum $k$ will exit the horizon when $k = a(t) H(t)$,
\begin{align}
n_s -1 = \frac{d {\rm ~log}~ P_{\zeta}}{d ~{\rm log}~ k}  = \frac{1}{H P_{\zeta}} \left(\frac{d}{dt} P_{\zeta} \right) = - 6 \epsilon - \frac{2 \lambda_{\sigma }\sigma^2}{H^2},
\label{eq:nsfin}
\end{align}
using Eq.~\eqref{eq:pzetafinquart}, the definition $\epsilon= - \frac{\dot{H}}{H^2}$, and the slow-roll equation $3 H \dot{\sigma} = - V_{\sigma}$.
These expressions for the power spectrum and spectral index constrain the curvaton's quartic self-coupling $\lambda_{\sigma}$ and initial field value, $\sigma_*$, in terms of $n_s^*$ and $A_s^*$, and $\epsilon$. We find
\begin{align}
\sigma_* = \frac{3 H_*}{\pi \sqrt{A_s^*}} \frac{1}{(1 - n_s^* - 2 \epsilon)}
\label{eq:sigstar}
\end{align} 
and 
 \begin{align}
\lambda_{\sigma} = \frac{\pi^2 A_s^*}{18}  (1 - n_s^* - 2 \epsilon)^3 \, .
\label{eq:lambdasigfull}
\end{align} 

To good approximation, especially in the case of small field inflation, $\epsilon \sim 0$. As a result, we can express $\lambda_{\sigma}$ as a function of only $A_s^*$ and $n_s^*$. Inserting the Planck collaboration's $1 \sigma$ preferred values for $A_s^*$ and $n_s^*$,
\begin{align}
1.9 \times 10^{-14} \leq \lambda_{\sigma} \leq 6.9 \times 10^{-14}
\label{eq:lambdasigbound}
\end{align} 
Note that this prediction for $\lambda_{\sigma}$ is independent of $\Lambda$, $N_*$, and the inflaton's potential.

\begin{figure}[h]
\centering
\includegraphics[width=.6\textwidth]{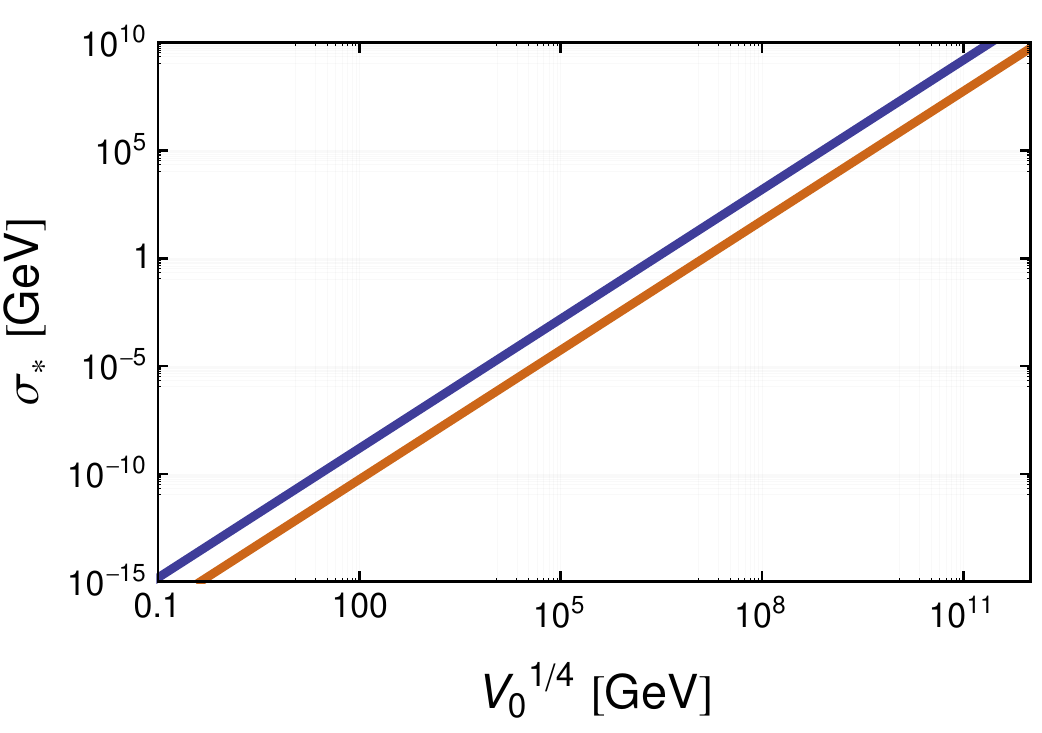}
\caption{The upper blue line shows what values of the initial curvaton field value ($\sigma_*$) are consistent with Planck observations of $n_s^*$ and $A_s^*$, for the quartic hilltop curvaton potential, Eq.~\eqref{eq:curvatonquartic}. As explained in the text, the required value of $\sigma_*$ is independent of $\Lambda$ and $N_*$. Allowing variations corresponding to Planck's $1 \sigma$ bounds on $n_s^*$ and $A_s^*$ has little effect in this plot, generating a thickness less than that of the blue line. The orange line is a lower bound for $\sigma_*$, such that $\sigma$ is slowly rolling during inflation rather than dominated by quantum fluctuations -- the quartic curvaton satisfies this requirement.}
\label{fig:sigmastar}
\end{figure}
  
Turning to Eq.~\eqref{eq:sigstar}, and inserting the relation $H_*^2 = \frac{V_{0}}{3 M_{\rm p}^2}$, one finds that $\sigma_*$ depends only on $V_{0 }^{1/4}$ and is independent of $N_*$ or $\Lambda$. Using this, in Figure \ref{fig:sigmastar} $\sigma_*$ is plotted as a function of $V_{0 }^{1/4}$ with a blue line. Note that setting the thickness of the blue line to coincide with Planck's $1 \sigma$ constraints on $n_s$ and $A_s$ would generate a line too thin to be seen, so we plot this line with a machine thickness. Figure \ref{fig:sigmastar} also shows an orange line, which is the minimum $\sigma_*$ such that $\sigma$ will be slowly rolling during inflation, rather than in a regime dominated by quantum fluctuations. The blue line, corresponding to $\sigma_*$ is always above the orange line, and so the curvaton will be slowly rolling in parameter space matching primordial perturbations observed on the CMB. This demonstrates that it was appropriate to use the slow-roll approximation in our treatment of the curvaton. To plot the orange line, we have used the curvaton's equation of motion, along with the standard requirement that the field distance the curvaton rolls in one Hubble time ($\Delta \sigma_* \approx  \frac{\partial_{\sigma} V}{3 H_*^2}$) is larger than its quantum fluctuations in de Sitter space ($\delta \sigma \sim \frac{H_*}{2 \pi}$),
\begin{align}
\sigma_*^3 \geq \frac{3 H_*^3}{2 \pi \lambda_\sigma}.
\end{align}

Next, to validate our use of a perturbative expansion in computing curvaton primordial perturbations, we must ensure the initial field value $\sigma_*$ is greater than de Sitter-induced variations to the curvaton's field value ($\delta \sigma$) during inflation. In other words, the fluctuations of the curvaton during inflation should be small, compared to the initial field value of the curvaton, $\delta \sigma \ll \sigma_*$. So long as this is satisfied, the perturbative formulae used to calculate curvaton primordial perturbations will be valid. 
$\sigma_*$ was calculated above so that it would produce the observed $A_s$ and $n_s$,
\begin{align}
\sigma_* = \frac{3 H_*}{\pi \sqrt{A_s} (1-n_s)}
\end{align}
which implies
\begin{align}
\frac{\sigma_*}{H_*} = 6.408 \times 10^5 \, .
\end{align}
Thus, for the curvaton under consideration, the perturbative regime holds.

Some comments are in order about non-Gaussian perturbations from the hilltop curvaton studied here. The non-Gaussian perturbations produced by a hilltop curvaton are characteristically small enough to lie within Planck's $1 \sigma$ bound on the lowest-order non-Gaussian parameter, $f_{NL} = 2.7 \pm 5.4$ \cite{Ade:2015ava}, but exact conclusions depend upon the curvaton's evolution after inflation, prior to its decay \cite{Sasaki:2006kq,Kawasaki:2011pd}. It would be interesting to further analyze the non-Gaussian curvaton signatures produced by various reheating scenarios, and relate them to findings in this study, for the quartic hilltop curvaton and other low scale models. This would allow for verification of a putative inflaton-curvaton pair found at a collider, with a future measurement of non-Gaussianity, assuming a vigorous 21-cm cosmological program that permits the detection of such small non-Gaussianities \cite{Cooray:2006km,Pritchard:2011xb,Munoz:2015eqa}.

So far, the self coupling ($\lambda_\sigma$) and initial field value ($\sigma_*$) have been determined, for a curvaton with a quartic hilltop potential, to reproduce the primordial perturbations observed on the CMB. There are some additional stipulations on curvaton parameter space. Section \ref{subsec:curvatoninflation} addresses the requirement that the curvaton \emph{not} produce a substantial second wave of inflation. Section \ref{subsec:curvatonreheat} discusses the curvaton energy density, which must become the predominant energy density of the universe sometime after the inflaton decays. This places a bound on the duration and average equation of state during reheating.

\subsection{Limiting curvaton-induced inflation}
\label{subsec:curvatoninflation}
One constraint on curvaton parameter space arises from the assumption that the curvaton does not generate a second period of inflation after the end of $\phi$-driven inflation. At the end of inflation, by construction the inflaton dominates the energy density of the universe. Then in a curvaton cosmology, the inflaton decays into lighter fields more rapidly than the curvaton. After the inflaton has decayed, the curvaton energy density $V(\sigma_{end})$ will at first remain nearly constant, as the curvaton is slowly rolling down its potential during radiation-dominated expansion. More precisely, it can be verified using the slow-roll equations, that the requisite flatness of the quartic curvaton potential used in this study, results in the curvaton slowly rolling at least until the energy density of the universe dilutes to $\sim V_{0\sigma}$. Once the energy density of the universe dilutes to $\sim V_{0\sigma}$, the curvaton begins oscillating in its potential. However, if at this time the curvaton is still slowly rolling, a short period of curvaton-driven inflation can occur.  

To rule out a substantial period of curvaton-driven inflation, it is sufficient to estimate the amount of curvaton-driven inflation that would result if the curvaton's energy density became predominant immediately at the end of $\phi$-driven inflation. The energy density in the curvaton field at the end of inflation is $\sim V_{0 \sigma} \simeq \lambda^3_{\sigma} \Lambda^4 /432$. Here, $V_{0 \sigma}$ has been calculated by using the quartic curvaton potential given above. In other words,  we wish to check whether the subdominant portion of the total vacuum energy, $V_{0 \sigma}$, will be a substantial source of inflation as $\sigma$ rolls to its minimum. This will depend on how slowly $\sigma$ rolls to its minimum.

First, it is necessary to calculate the field value of the curvaton at the end of inflation, $\sigma_{end}$. We employ the slow roll formula to find how far the curvaton rolls during inflation,
$
3 H \dot{\sigma} \approx - \partial_{\sigma} V,
$ which integrates to
\begin{align}
\sigma_{end} = \sqrt{\frac{1}{ \frac{2 \lambda_{\sigma}}{3 H} (t_* - t_{end}) + \frac{1}{\sigma_*^2}  }} \, .
\end{align}
This field value can be re-written in terms of $N_*$ using that $N = \int H dt$, $i.e.$ $\Delta N \approx H \Delta t$.
\begin{align}
\sigma_{end} = \sqrt{\frac{1}{ -\frac{2 \lambda_{\sigma} N_*}{3 H^2} + \frac{1}{\sigma_*^2}  }}
\end{align}
The curvaton potential matches that of the inflaton, so the same formula, Eq.~\eqref{eq:Nefforsimpmodel}, gives the number of efolds for curvaton-driven inflation.
Requiring curvaton-driven inflation last less than one efold,
\begin{align}
\frac{\pi^6 A_s^3 (1-n_s)^8 \Lambda^4}{2^7 3^8 V_0} \left(1 - \frac{N_* (1-n_s)}{3} \right) \leq 1 \, .
\end{align}
Using Planck's central values for $n_s$ and $A_s$, this in turn yields,
\begin{align}
V_0^{\frac{1}{4}} \geq 5.97 \times 10^{-11} \Lambda \left(1 - 0.01 N_* \right)^{\frac{1}{4}},
\label{eq:boundcurvinfl}
\end{align}
which is plotted in Figure \ref{fig:V0LamCurv}. While it might be possible to consider curvaton inflation that lasts for up to $\sim 15$ efolds, before the ``curvaton" would be an inflaton producing (disallowed) perturbations on CMB scales, the cosmological consistency conditions in Section \ref{sec:reheatingconstraints}, which are accurate to within about an efold, would have to be recalculated for each point in this parameter space. This would greatly complicate the treatment of curvaton and inflaton parameter space presented hereafter, without qualitatively changing results, because the bound of Eq.~\eqref{eq:boundcurvinfl} would change by less than a factor of two.

\subsection{Curvaton during reheating}
\label{subsec:curvatonreheat}
For a curvaton cosmology, the period of reheating lasts until the curvaton decays. At the time of curvaton decay, the curvaton's energy density must have grown substantially larger than that of the inflaton (more precisely, the inflaton's decay products), so that isocurvature perturbations are minimized, in accord with Planck's 2015 bound on isocurvature modes \cite{Ade:2015lrj}. Planck's bound on the leftover inflaton energy density is $\beta_{iso} \leq 0.0013$ (using data from TT, TE, EE + lowP), which is equivalent to requiring that the curvaton comprise 99.1\% of the total energy density of the universe before the end of reheating,
\begin{align}
\frac{\rho_{\phi}}{\rho_{tot}} \Big|_{re} \leq 0.0089,
\label{eq:planckisobound}
\end{align}
where $\rho_{tot} \equiv \rho_\sigma +\rho_\phi$, and here $|_{re}$ indicates the end of reheating. Of course, each of $\rho_{\phi}$, $\rho_\sigma$ indicate the summed energy density of $\phi,~\sigma,$ and their respective decay products. Assuming the inflaton $\phi$ decays promptly after inflation into radiation, the energy density of $\phi$ in the universe at the end of reheating will be
\begin{align}
\rho_{\phi,re} = \rho_{\phi,ei} e^{-4 N_{re}},
\label{eq:rhophi}
\end{align}
where $|_{ei}$ indicates the end of inflation, and $N_{re}$ is the number of efolds during reheating, as in Section \ref{sec:reheatingconstraints}. The total energy density on the other hand is given by,
\begin{align}
\rho_{tot,re}  = \rho_{tot,ei} e^{-3 N_{re}(1 +   w_{re})},
\label{eq:rhotot}
\end{align}
where again $w_{re}$ is defined as the average equation of state during reheating. The total energy density at the end of inflation is approximately the inflaton's energy density. Thus, combining Eqs.~\eqref{eq:planckisobound}-\eqref{eq:rhotot}, sets a requirement on the number of efolds during reheating,
\begin{align}
N_{re} \geq \frac{4.72}{1-3 w_{re}},
\label{eq:Nrecurv} 
\end{align}
where here the numerator is simply the natural logarithm of Eq.~\eqref{eq:planckisobound}.

\begin{figure}[t!]
\centering
    \includegraphics[width=.6\textwidth]{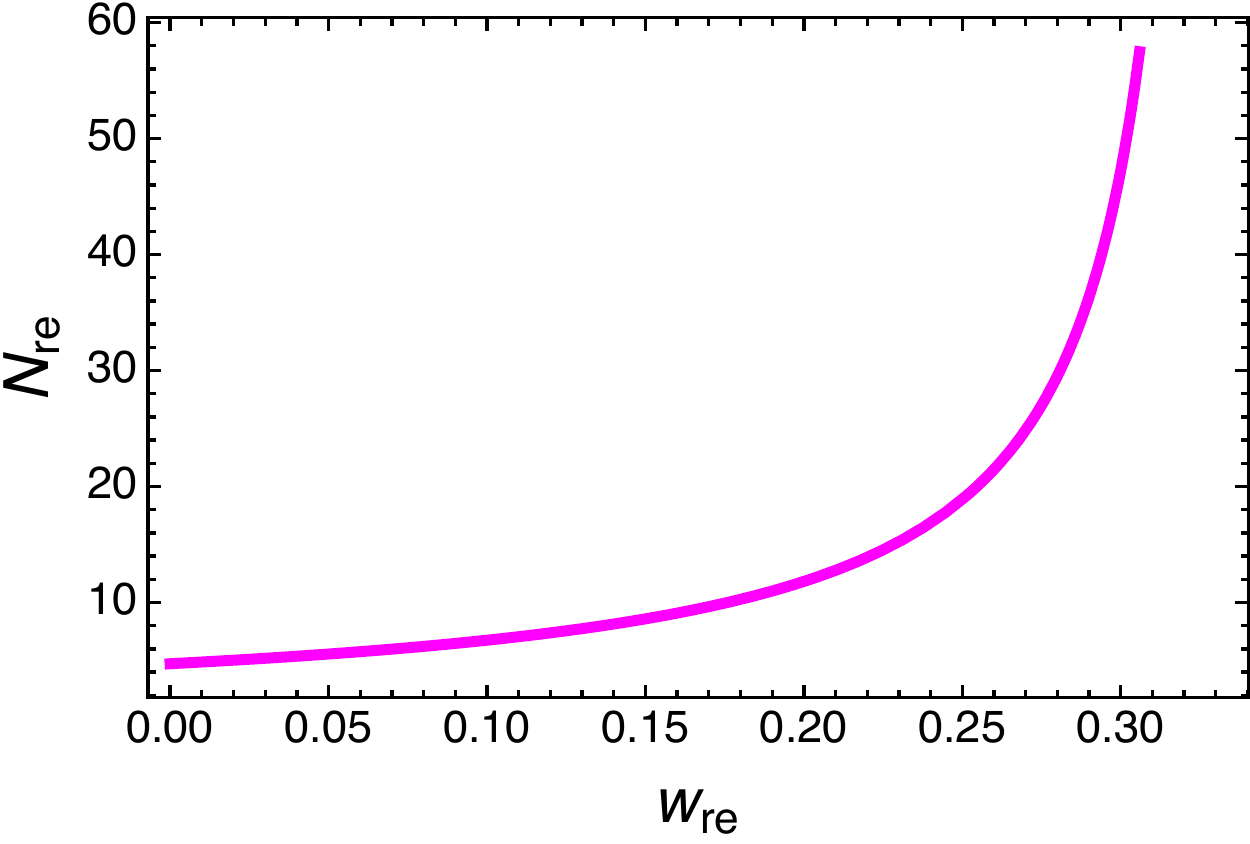}
    \caption{The plot shows the minimum number of efolds during reheating, $N_{re}$, as a function of $w_{re}$, the average equation of state during reheating, using Planck's constraint on isocurvature perturbations \cite{Ade:2015lrj}, $\beta_{iso} \leq 0.0013$. The region below the curve is excluded. A stiffer equation of state during reheating ($w_{re} \gg 0$) implies that the period of reheating must last longer, because it will take the curvaton energy density longer to become the predominant energy density in the universe. }
   \label{fig:cur}
\end{figure}

Figure \ref{fig:cur} plots this lower bound on $N_{re}$ as a function of $w_{re}$. The closer $w_{re}$ is to a radiation-like equation of state ($w_{re} \sim \frac{1}{3}$), the longer reheating must last so that the curvaton grows to dominate the universe's energy density.

Next we constrain $w_{re}$ in terms of $\Lambda$ and $V_0$. Note that since $w_{re}$ is the average equation of state during reheating,
\begin{align}
w_{re} = \frac{1}{N_{re}} \int w(N) \, d N
\end{align}
where
\begin{align}
w \equiv \frac{p_{tot}}{\rho_{tot}} = \frac{w_\phi \rho_\phi + w_{\sigma} \rho_{\sigma}}{\rho_\phi + \rho_{\sigma} }
\label{eq:wredef}
\end{align}

\begin{figure}[t!]
\centering
    \includegraphics[width=.49\textwidth]{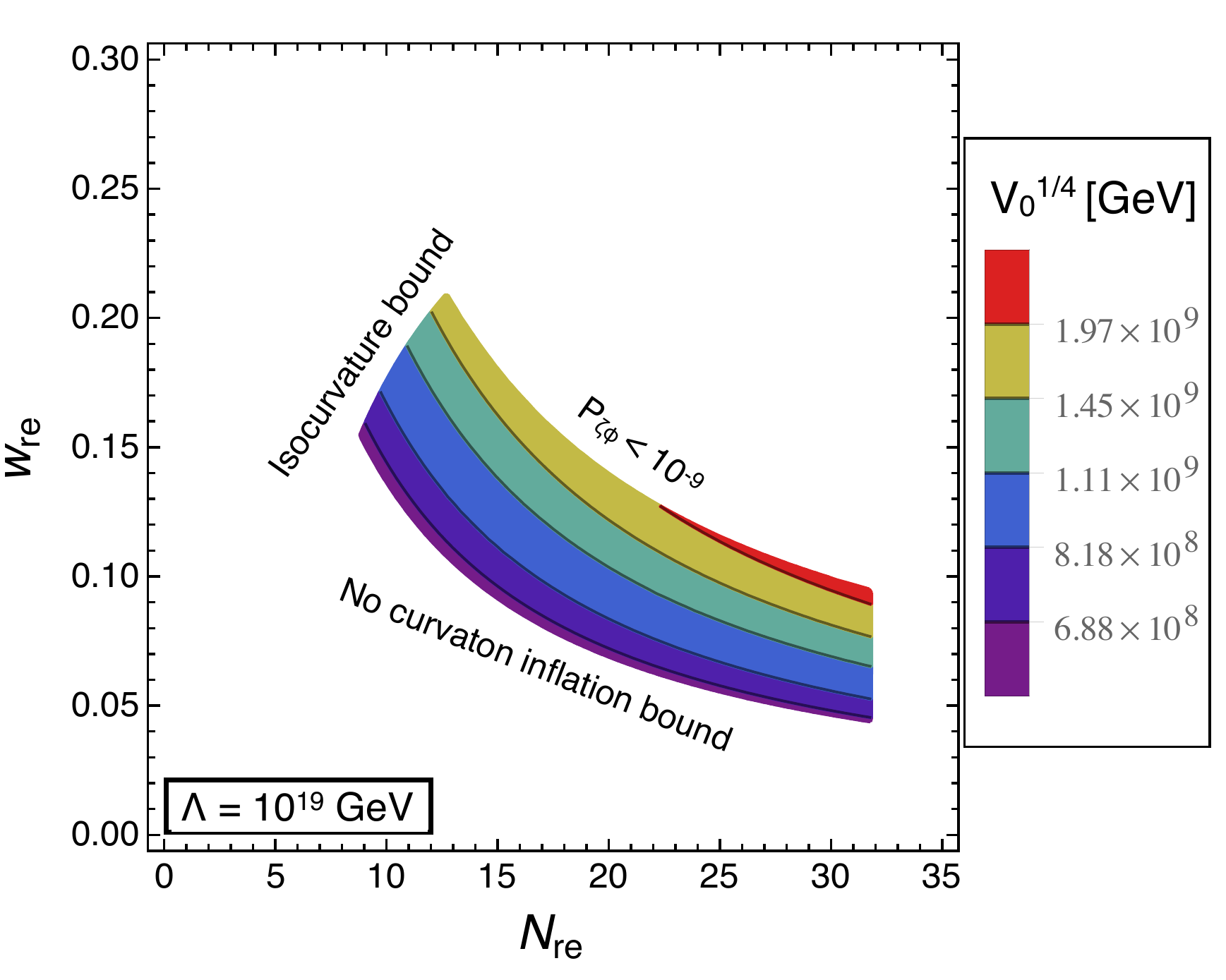}~
        \includegraphics[width=.49\textwidth]{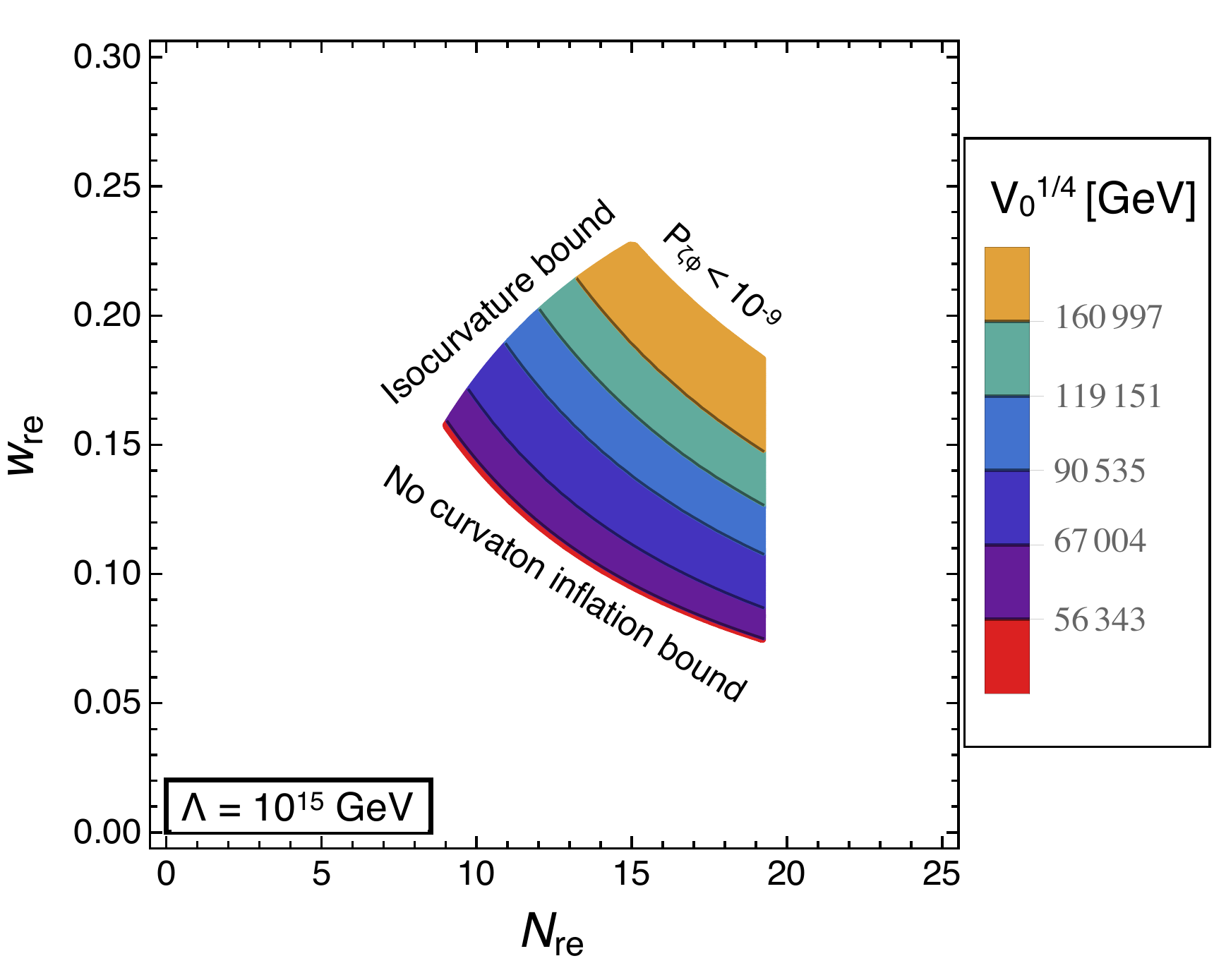}
    \caption{The figure shows the relationship between the energy density during inflation ($V_0$), the average equation of state during reheating ($w_{re}$), and the number of efolds during reheating ($N_{re}$), for a low scale quartic inflaton paired with a quartic curvaton and a given value of the cutoff $\Lambda$, as detailed in this section. The cosmological consistency conditions outlined in Section \ref{sec:reheatingconstraints} uniquely determine the number of efolds ($N_*$) for each point in parameter space. The parameter space is bounded at each edge by various factors. Top: the inflaton must not overproduce perturbations, Eq.~\eqref{eq:V0Lambound}. Bottom: the curvaton field must not produce more than an efold of inflation, Eq.~\eqref{eq:boundcurvinfl}. Left side: reheating must last long enough that the curvaton comes to dominate (constitute $\ge 99.1 \%$) of the energy density of the universe, Eqs.~\eqref{eq:Nrecurv} and \eqref{eq:wreV0}.  Note that the ``left side" bound assumes the curvaton equation of state after inflation is immediately matter-like, $i.e.$ $w_{\sigma,ei} \simeq 0$ -- relaxing this assumption would allow for a broader range of parameters. The viable parameter space is shown for cutoffs $\Lambda =10^{19}$ and $10^{15}$ GeV, as indicated.}
    \label{fig:rainbow}
\end{figure}

If we now assume that $w_{\sigma} \approx 0$, in other words that the curvaton behaves like matter during reheating, then $w_{\sigma} \rho_{\sigma} \simeq 0$. By assumption, we know that $\rho_{\phi, ei} \simeq V_{0}$. The energy density of the curvaton after inflation, $\rho_{\sigma , ei} \simeq V_{0 \sigma} = \lambda^3_{\sigma} \Lambda^4 /432$, where the final equality follows from the form of the curvaton potential, explained around Eq.~\eqref{eq:curvatonquartic}. 
Substituting these into Eq.~\eqref{eq:wredef} and integrating,
\begin{align}
\label{eq:wreV0}
 \frac{\lambda^3_{\sigma} \Lambda^4}{2^4 3^3 V_0} = \frac{1 - e^{N_{re} (3 w_{re} -1)}}{e^{3 N_{re} w_{re}} - 1} \, .
\end{align}

Combining equations \eqref{eq:Nrecurv} and \eqref{eq:wreV0} yields the left side bound on parameter space shown in Figure \ref{fig:rainbow}. Note as $w_{re} \rightarrow \frac{1}{3}$ the constraint becomes stronger. As in Figure \ref{fig:cur}, this is expected, since if $w_{re} = \frac{1}{3}$, then the equation of state of the curvaton would also be $w_\sigma=\frac{1}{3}$, and the curvaton energy density would not grow larger than the inflaton's energy density.

\subsection{Counting efolds for a quartic inflaton-curvaton pair}

\begin{figure}[h!]
\centering
    \includegraphics[width=.66\textwidth]{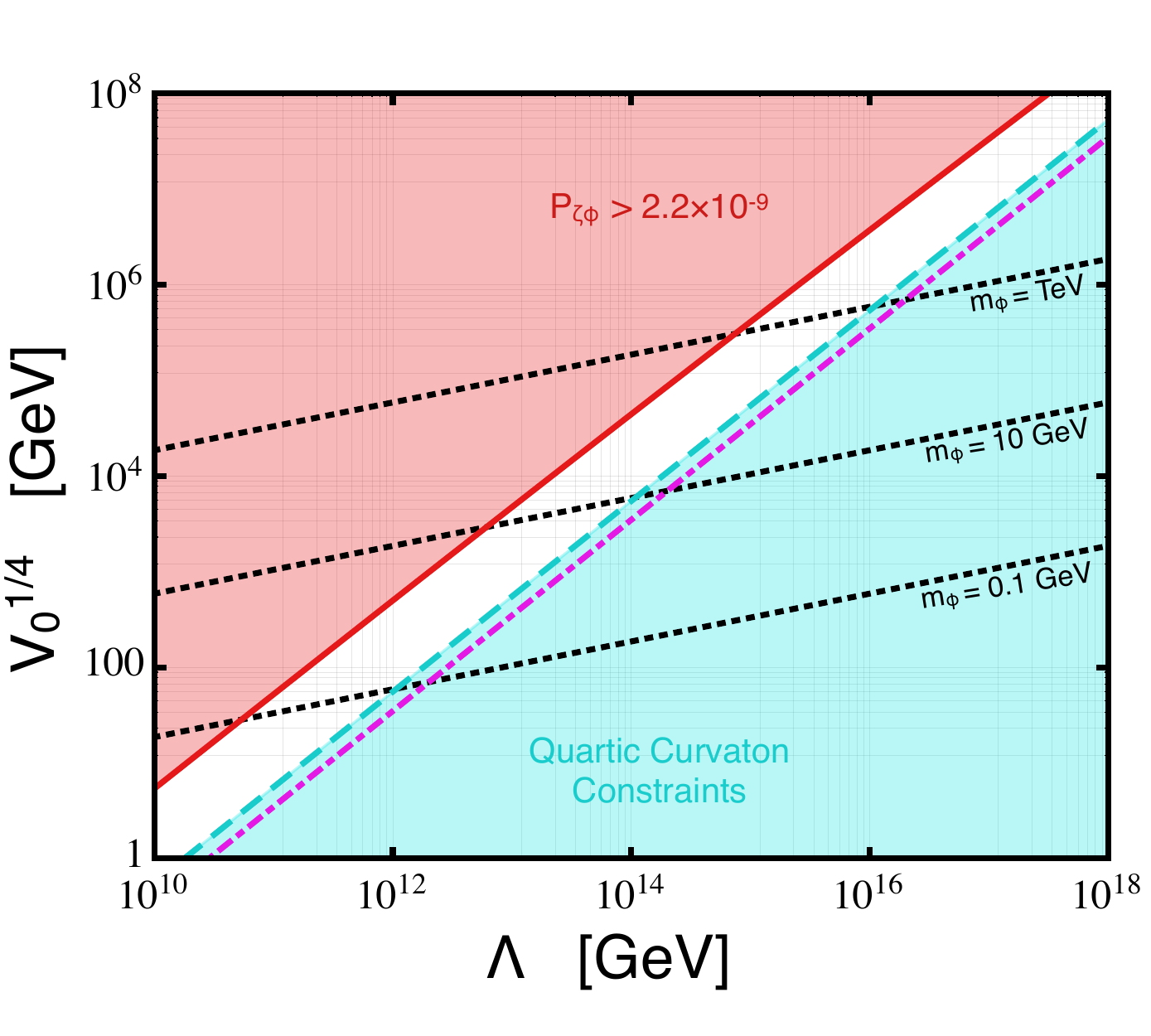}
    \caption{Similar to Figure \ref{fig:V0Lambda}, but with the addition of constraints from requiring a consistent curvaton cosmology. The top red line is again an upper bound, from requiring that the inflaton not over-produce perturbations, Eq.~\eqref{eq:V0Lambound}. For this bound, the number of efolds has been set with Eq.~\eqref{eq:nsmin}, which was derived using results in Section \ref{sec:reheatingconstraints}.  As in Figure \ref{fig:V0Lambda}, the dotted black lines show the implied mass of the inflaton.  All the other lines are lower bounds on viable parameter space. The blue dashed line marks the constraint that $\sigma$ not generate a second wave of inflation which lasts longer than an efold, Eq.~\eqref{eq:boundcurvinfl}, using Eq.~\eqref{eq:nsmax} to set the maximum number of efolds. The region above the pink dotted-dashed line indicates parameter space where $V_0 > 10 V_{0\sigma}$.}
    \label{fig:V0LamCurv}
\end{figure}

The constraints given by Eqs.~\eqref{eq:V0Lambound} and \eqref{eq:boundcurvinfl}, along with a line demarcating where the height of the inflaton potential exceeds the curvaton potential, $V_0 = 10 V_{0\sigma}$, are displayed in Figure \ref{fig:V0LamCurv}.  In this plot, the upper and lower bounds on the energy density during inflation ($V_0$) given in Eqs.~\eqref{eq:V0Lambound} and \eqref{eq:boundcurvinfl} have been sharpened by using the cosmological consistency equations, \eqref{eq:Nre} and \eqref{eq:Tre}. The line demarcating $V_0 = 10 V_{0\sigma}$, demonstrates that the inflaton energy density exceeds the curvaton energy density in all un-excluded parameter space.  Therefore it is correct to assume that $V_0 \gg V_{0\sigma}$ after inflation. This will be important for relating curvaton and inflaton decay widths in Section \ref{sec:maps}. 

Precisely, for fixed $\Lambda$, test values of $N_*$ can be specified, and a maximum and minimum $N_*$ value can be converged upon by requiring a solution to Eqs.~\eqref{eq:Nre} and \eqref{eq:Tre}, for a reasonable equation of state during reheating, $w_{re} =[0,\frac{1}{3}]$.  These minimum and maximum $N_*$ values can be used in Eqs.~\eqref{eq:V0Lambound} and \eqref{eq:boundcurvinfl}.  To illustrate the iterative procedure for computing $N_{*,\rm max},N_{*,\rm min}$, note that in Figure \ref{fig:reheatcons1}, where the cutoff $\Lambda =10^{19}~{\rm GeV}$ has been specified, the maximum number of efolds allowed by both the inflaton perturbation bound, Eq.~\eqref{eq:V0Lambound}, and the requirement $w_{re} = [0,\frac{1}{3}]$, is $N_{*,\rm max} \lesssim 41$.  By the same reasoning, for $\Lambda =10^{19}$ GeV, the bound on curvaton-driven inflation  Eq.~\eqref{eq:boundcurvinfl}, implies a minimum number of efolds, $N_{*,\rm min} \gtrsim 32$.  To see the provenance of this lower bound, note that $V_0^{1/4} \gtrsim 10^{10}$ GeV is required by Eq.~\eqref{eq:boundcurvinfl} for $\Lambda = 10^{19}$ GeV, and compare the allowed $V_0^{1/4}$ values in the $w_{re} =[0,\frac{1}{3}]$ region, of the top right and lower left panels of Figure \ref{fig:reheatcons1}. Given that there is a minimum reheating temperature $T_{re} > 4.7~$MeV, there is evidently a lower bound on efolds, $N_{*,\rm min}\sim 30-35$, if $w_{re} =[0,\frac{1}{3}]$.  

Put another way, visual inspection of Eqs.~\eqref{eq:V0Lambound} and \eqref{eq:boundcurvinfl} reveals that they have a weak dependence on $N_*$, for the overall allowed range $N_* \simeq 20-40$.  For a given value of $\Lambda$ or $V_0$, one can iteratively specify test $N_*$ values and use Eqs.~\eqref{eq:Nre} and \eqref{eq:Tre} to converge upon $N_{*,\rm max},N_{*,\rm min}$.  Performing this iterative computation on a representative sample of $\Lambda$ values, the following relations can be derived with a numerical fit,
\begin{align}
N_{*,\rm max} \simeq 21.5 + 0.023 ~\left({\rm log} \left[\frac{V_0^{1/4}}{\rm GeV} \right] \right)^2
\label{eq:nsmin}
\end{align}
and
\begin{align}
N_{*,\rm min} \simeq 19.0 + 0.986~  {\rm log} \left[\frac{V_0^{1/4}}{\rm GeV} \right] 
\label{eq:nsmax}
\end{align}

Both the maximum and minimum number of efolds increase with $V_0$, because a higher energy density at the end of inflation implies a lengthier expansion of the comoving horizon during reheating and the radiation-dominated epoch.  In Figure \ref{fig:V0LamCurv}, these expressions for the minimum and maximum number of efolds have been used to produce more precise bounds on $V_0$ as a function of $\Lambda$, specifically the bound on the inflaton over-producing perturbations, Eq.~\eqref{eq:V0Lambound} and the bound on the curvaton producing a second epoch of inflation, Eq.~\eqref{eq:boundcurvinfl}.

\section{Predicting the curvaton from the inflaton and vice-versa}
\label{sec:maps}

This section shows how results in Sections \ref{sec:simplequarticmodel}--\ref{sec:curvaton} can be used to predict the mass of the inflaton from the mass of the curvaton, and vice-versa, to within about an order of magnitude. Also, it will be demonstrated that the inflaton decay width sets an upper bound on the decay width of the curvaton, and the curvaton's decay width sets a lower bound on that of the inflaton.  In Section \ref{sec:inflatonhiggsportal}, these relations will be used to relate low scale inflaton-curvaton pairs, coupled to the Standard Model through a Higgs portal.

In Section \ref{sec:curvaton}, the average equation of state ($w_{re}$) and number of efolds ($N_{re}$) during reheating were employed to parameterize the collective cosmological behavior of the inflaton and curvaton. This provided a set of viable cosmological histories, assuming the inflaton decayed instantaneously at the end of inflation, as illustrated in Figures \ref{fig:rainbow} and \ref{fig:V0LamCurv}. Importantly, this analysis also bounded the number of efolds of inflation for reasonable reheating scenarios, Eqs.~\eqref{eq:nsmin} and \eqref{eq:nsmax}. In this section, these results will allow us to directly address the decay widths of the inflaton and curvaton, $\Gamma_\phi$ and $\Gamma_\sigma$, without having to precisely specify $w_{re}$ or $N_{re}$. 

Before exploring parametric maps between inflaton and curvaton parameter space, we can first calculate a general upper bound on the decay width of the curvaton ($\Gamma_{\sigma}$) to Standard Model particles. After inflation, the curvaton must not decay for a long enough time period that it comes to dominate the energy density of the universe. An upper bound on the curvaton decay width can be derived by noting that to good approximation, while it is oscillating in its potential, the curvaton field dilutes like matter in an expanding universe. 

As explored in Section \ref{sec:curvaton}, the curvaton will slowly roll, and its energy density $V_{0\sigma}$ will remain approximately constant, until the inflaton's energy density (initially $\sim V_0$) has diluted enough that $\rho_\phi \sim V_{0\sigma}$. At this time, the curvaton begins oscillating in its potential and diluting like matter ($\propto a^{-3}$), while the inflaton's radiation-like energy density dilutes as $a^{-4}$. Thereafter, once the universe has expanded further by a factor of $\Delta a \gtrsim 100$, the curvaton energy density will exceed the inflaton's energy density by a factor of $\gtrsim 100$ -- as required by the Planck bound on isocurvature fluctuations ($\frac{\rho_\phi}{\rho_{tot}} < 0.0089$) given in Section \ref{sec:curvaton}. Using the instantaneous decay approximation, the total decay width of the curvaton will be approximately equal to the Hubble constant when the curvaton decays $\Gamma_\sigma \sim H_\sigma$. Thus, using the relation $3 H^2 = \rho/M_{p}^2$, that the energy density of the universe when the curvaton begins oscillating ($\rho_\phi \sim V_{0 \sigma}$) will dilute as $a^{-3}$, and that $\Delta a \gtrsim 100$, the maximum conceivable curvaton decay width consistent with a quartic hilltop inflaton cosmology is
\begin{align}
\Gamma_{\sigma} \leq  \frac{\sqrt{V_{0\sigma}}}{10^3 \sqrt{3} M_{\rm p}} \simeq 10^{-8} \left(\frac{m_\phi}{{\rm GeV}} \right)^2~{\rm GeV}~ ,
\label{eq:Gammasig1}
\end{align}
where this expression was re-phrased in terms of the inflaton mass by incorporating Eqs.~\eqref{eq:mphi},~\eqref{eq:lambdasigbound},~\eqref{eq:boundcurvinfl}, $V_{0\sigma} = \lambda_\sigma^3 \Lambda^4 /432$, the limiting case of $N_* =40$ and $\lambda_\sigma = 6.9 \times 10^{-14}$, and that the Hubble constant $H = \sqrt{V_{0 \sigma}/3M_{\rm p}^2}$ scales with $a^{-3/2}$ during matter-dominated expansion. 

An upper bound on the curvaton decay width is given above; a trivial lower bound on the curvaton decay width arises from requiring that the curvaton decay before the onset of BBN, $\Gamma_\sigma \gtrsim 10^{-23}~{\rm GeV}$ ($T_{re} > 4.7~{\rm MeV}$). Note that these bounds on the curvaton decay width hold, irrespective of the curvaton model assumed. Therefore, any terrestrial experiment sensitive to new scalar states with decay widths ranging from $10^{-2}$ -- $10^{-23}~{\rm GeV}$ are potentially sensitive to a low scale curvaton. In more detail, Section \ref{sec:inflatonhiggsportal} will show that upcoming searches of Higgs portal parameter space, probe the desired range of decay widths, and importantly within a parameter space that does not spoil the flatness of the hilltop inflaton and curvaton potentials.

\subsection*{Mapping a quartic inflaton to a quartic curvaton}

Because the inflaton and curvaton have potentials of the same form, the formula for the mass of the curvaton at its minimum matches that of the inflaton, Eq.~\eqref{eq:mphi}, with the replacement $V_0 \rightarrow V_{0 \sigma}$. As explained in the preceding section, the height of the curvaton potential is given by $V_{0\sigma} = \lambda^3_{\sigma} \Lambda^4 /432$, so altogether, 
\begin{align}
m_{\sigma} = \frac{\lambda_{\sigma} \Lambda}{\sqrt{3}}~.
\label{eq:msigma}
\end{align}
With $\Lambda$ specified, the curvaton mass at its minimum can be determined within observational bounds, since Planck's $1 \sigma$ bound on $n_s$ and $A_s$ restrict the curvaton quartic coupling, $1.9 \times 10^{-14} \leq \lambda_{\sigma} \leq 6.9 \times 10^{-14}$ (see Section \ref{sec:curvaton}). Furthermore, inspecting Figure \ref{fig:V0LamCurv}, it is clear that if an inflaton of mass $m_\phi$ is discovered, $\Lambda$ and as a consequence $m_\sigma$, will be restricted to within about an order of magnitude. Combining Eqs.~\eqref{eq:mphi}, \eqref{eq:V0Lambound}, and \eqref{eq:msigma}, we find that requiring the inflaton not over-produce perturbations during inflation results in
\begin{align}
m_{\sigma} \gtrsim 0.051~m_\phi \left( \frac{N_*}{25} \right)^3 \left( \frac{4 \times 10^{-10}}{P_{\rm \zeta \phi}} \right),
\label{eq:sigmasslower}
\end{align}
where here we take the lower Planck $1 \sigma$ preferred value $\lambda_\sigma = 1.9 \times 10^{-14}$, and normalize so that the inflaton perturbations are one-fifth as large as those observed, to avoid substantially altering curvaton perturbations \cite{Langlois:2004nn,Sloth:2005yx}.
Similarly, combining Eqs.~\eqref{eq:mphi}, \eqref{eq:msigma}, and \eqref{eq:boundcurvinfl}, which requires that the curvaton not over-inflate the universe,
\begin{align}
m_{\sigma} \lesssim 0.39~m_\phi \left(1-0.01 N_* \right)^{-1/3},
\label{eq:sigmassupper}
\end{align}
where this expression takes the limiting case of $\lambda_\sigma = 6.9 \times 10^{-14}$. 

An upper bound on the curvaton decay width can be set directly using the inflaton's decay width, by noting (as derived at the outset of this section) that the Hubble constant of the universe must dilute by $\sim 100^{-3/2}$ before curvaton decay,
\begin{align}
\Gamma_{\sigma} \leq \frac{\Gamma_{\phi}}{10^3} .
\label{eq:absGsigma}
\end{align}
For the inflaton and curvaton Higgs portal parameter space considered in the next section, this bound on the curvaton decay width is stronger than that of Eq.~\eqref{eq:Gammasig1}. Some example inflaton masses and decay widths are listed in Table \ref{tab:paramaps}, along with a predicted range of values for corresponding curvaton masses and decay widths. In this Table, Eqs.~\eqref{eq:nsmax} and \eqref{eq:nsmin} are employed to determine the number of efolds in Eqs.~\eqref{eq:sigmasslower} and \eqref{eq:sigmassupper}. 

\begin{table}[t]
   \centering
   \begin{tabular}{@{} lcccccc @{}}
      \toprule 
      $m_\phi$ (GeV)   & $\Gamma_\phi$ (GeV) & $\Lambda$ (GeV)  &  $V_0^{1/4}$ (GeV) & $N_*$   &$ m_\sigma$ (GeV) & $ \Gamma_\sigma$ (GeV) \\
      \midrule
       0.3       &  $10^{-20}$   & (0.9 -- 3)$ \times 10^{12}$ & 130--180  &  22--26   & 0.01--0.13           &      $\leq 10^{-23}$        \\
       4         &  $10^{-18}$   & (1 -- 4)$ \times 10^{13}$     & 1700--2400  &  23--28  &  0.18--1.7            &      $\leq 10^{-21}$        \\
       100       &  $10^{-5}$   &  (0.4 -- 1)$ \times 10^{15}$     & (4.6 -- 6.1)$ \times 10^4$   &  24--31   &    4.5--44            &      $\leq 10^{-8}$        \\
      \bottomrule
   \end{tabular}
    \begin{tabular}{@{} lcccccc @{}}
      \toprule 
      $m_\sigma$ (GeV)   & $\Gamma_\sigma$ (GeV) & $\Lambda$ (GeV)  &  $V_0^{1/4}$ (GeV)  & $N_*$   &    $ m_\phi$ (GeV) & $ \Gamma_\phi$ (GeV) \\
      \midrule
       0.3       &  $10^{-20}$  & (0.8 -- 3)$\times 10^{13}$  & 410--3300  & 22--27   &   0.7--8.2           &      $\geq 10^{-17}$        \\
       4        &  $10^{-18}$  & (1 -- 3.6)$\times 10^{14}$  & (0.5 -- 4.5)$\times 10^{4}$  & 23--29  &   9.3--103            &      $\geq 10^{-15}$       \\
       100      &  $10^{-5}$  &  (2.5 -- 9)$\times 10^{14}$  &  (1 -- 9)$\times 10^{5}$  & 25--32  &  230--2000            &      $\geq 10^{-2}$       \\
      \bottomrule
   \end{tabular}
   \caption{The implied quartic curvaton mass and decay width, $m_\sigma$ and $\Gamma_\sigma$, for given quartic inflaton mass and decay width, $m_\phi$ and $\Gamma_\phi$ (top table), and vice-versa (bottom table), using cosmological constraints detailed around Eqs.~\eqref{eq:sigmasslower}-\eqref{eq:absGphi}. The main factors setting the range of predicted values, are the requirements that the inflaton produce small perturbations, $P_{\rm \zeta \phi} < 4\times 10^{-10}$, that curvaton-induced inflation lasts for less than an efolding, that isocurvature perturbations are small, and that the curvaton produce the perturbations observed on the CMB (to within $1 \sigma$ of Planck's reported values for the power spectrum and spectral index).}
   \label{tab:paramaps}
\end{table}

\subsection*{Mapping a quartic curvaton to a quartic inflaton}

With a similar procedure, we can find upper and lower bounds on a small field quartic inflaton from measurements of a small field quartic curvaton. Using the Planck collaboration bound on $\lambda_\sigma$, Eq.~\eqref{eq:lambdasigbound}, along with Eqs.~\eqref{eq:mphi}, \eqref{eq:V0Lambound}, and \eqref{eq:msigma},
\begin{align}
m_\phi \lesssim 20 ~m_\sigma~  \left( \frac{25}{N_*} \right)^3 \left( \frac{P_{\rm \zeta \phi}}{4\times 10^{-10}} \right),
\label{eq:phimassupper}
\end{align}
where again we use that in the limiting case, $\lambda_\sigma = 1.9 \times 10^{-14}$, \eqref{eq:lambdasigbound}. Next we again combine Eqs.~\eqref{eq:mphi}, \eqref{eq:msigma}, and \eqref{eq:boundcurvinfl} to find a lower bound on the inflaton mass,
\begin{align}
m_\phi \gtrsim 2.6 ~m_\sigma \left(1-0.01 N_*\right)^{1/3}, 
\end{align}
where we have used the limiting value $\lambda_\sigma = 6.9 \times 10^{-14}$.

Finally, a lower bound on $\Gamma_\phi$ arises directly from Eq.~\eqref{eq:absGsigma},
\begin{align}
\Gamma_{\phi} \geq 10^3 \Gamma_{\sigma}~.
\label{eq:absGphi}
\end{align}
Some inflaton mass and decay width predicted from curvaton masses and decay widths are shown in Table \ref{tab:paramaps}, again using Eqs.~\eqref{eq:nsmin} and \eqref{eq:nsmax} to iteratively determine $N_*$.

Table \ref{tab:paramaps} also gives a range of $\Lambda$, $V_0^{1/4}$ and $N_*$ values predicted by the quartic inflaton-curvaton model for a given value of $m_\phi$ or $m_\sigma$. These ranges are visually apparent in Figure \ref{fig:V0LamCurv}, where a quartic inflaton with fixed $m_\phi$ is confined to a range of permitted $\Lambda$ and $V_0^{1/4}$ values. For a given $m_\phi$, the lower bound on $V_0^{1/4}$ (upper bound on $\Lambda$) follows directly from Eqs.~\eqref{eq:mphi} and \eqref{eq:V0Lambound}. The upper bound on $V_0^{1/4}$ (lower bound on $\Lambda$) follows directly from Eqs.~\eqref{eq:mphi} and \eqref{eq:boundcurvinfl}. The same relationships hold for $m_\sigma$, except using Eq.~\eqref{eq:msigma} instead of Eq.~\eqref{eq:mphi}. With a range of permitted $V_0^{1/4}$ values, a range of $N_*$ values can be obtained immediately from Eqs.~\eqref{eq:nsmin} and \eqref{eq:nsmax}.

\section{Higgs portals to low scale inflation}
\label{sec:inflatonhiggsportal}

A simple way for the inflaton and curvaton to couple to Standard Model particles in a renormalizable fashion (in this case allowing the inflaton or curvaton to dump its energy into a bath of Standard Model particles after the end of inflation) is through a Higgs portal operator \cite{Burgess:2000yq,Schabinger:2005ei,Patt:2006fw}. A scalar field coupled to the Higgs in this manner can be probed at the LHC \cite{Dawson:2009yx,Englert:2011yb,Fox:2011qc,Batell:2011pz,Curtin:2013fra,Freitas:2015hsa} and other low energy experiments, Refs.~\cite{Wilczek:1977zn,McKeen:2008gd,Batell:2009jf,Clarke:2013aya,Schmidt-Hoberg:2013hba,Dolan:2014ska,Alekhin:2015byh,Krnjaic:2015mbs}. In the case of low scale inflation, this section demonstrates that the inflaton-Higgs and curvaton-Higgs couplings can be small enough not to spoil the flatness of the inflaton or curvaton potentials through radiative corrections, while allowing for enough inflaton/curvaton-Higgs mixing to efficiently reheat the universe after inflation, all within parameter space accessible at upcoming low energy experiments.

A Higgs portal inflaton appearing at meson factories has been studied previously in the context of large field inflation, specifically for a scalar inflaton field non-minimally coupled to gravity \cite{Bezrukov:2009yw,Bezrukov:2013fca}. Non-minimally coupled inflation models rely on the inflaton potential becoming flat at large field values, as determined by the ultraviolet running of the inflaton's coupling to gravity. In the non-minimally coupled scenario of Refs.~\cite{Bezrukov:2009yw,Bezrukov:2013fca}, the observed spectrum of primordial perturbations restricts the inflaton mass to $m_{\varphi} \sim 0.27 -1.8 ~{\rm GeV}$. However, it is important to note that predictions in non-minimally coupled models of inflation, which by necessity have couplings that change substantially as they are RG-evolved to large field values, are sensitive to corrections from non-renormalizable operators -- and equivalently the unknown ultraviolet dynamics of the theory \cite{Burgess:2014lza}.

On the other hand, the low scale inflaton and curvaton sectors we consider here are very weakly coupled, both to themselves and to the Higgs boson. In spite of a miniscule coupling to the Higgs boson, the remainder of this section shows that the quartic hilltop inflaton (and its lighter curvaton partner) detailed in Sections \ref{sec:simplequarticmodel}--\ref{sec:maps} can be found through a Higgs portal at the LHC and other low energy experiments, over a broad mass range, $m_{\phi}, m_{\sigma} = {\rm MeV-TeV}$. The key point will be that the large VEV predicted for the inflaton and curvaton at their minima allows for sizable mixing with the Higgs, even though the actual Higgs portal coupling is tiny. 

In the treatment that follows, we will begin referring exclusively to the inflaton. Because the form of the quartic inflaton and curvaton potentials are identical, an identical treatment applies to the curvaton. For the parameter space we are interested in, the Higgs-inflaton and Higgs-curvaton couplings are each small enough, that the computation of a full $3 \times 3$ mixing matrix does not alter results.

We begin by extending the potential given in Eq.~\eqref{eq:simplequartic} to include the Higgs sector of the Standard Model, with the addition of a quartic inflaton-Higgs portal operator. Starting with $V( \Phi) = - \mu^2 \Phi^{\dagger} \Phi + \lambda_h (\Phi^{\dagger} \Phi)^2$, where $\Phi$ is the SM Higgs doublet. Prior to electroweak symmetry breaking, the potential is given by
\begin{align}
V_{\phi h} = V_0  -\frac{\lambda_\phi}{4} \phi^4 +\frac{\phi^6}{\Lambda^2}+\lambda_{\rm \phi h } |\Phi|^2 \phi^2 + \lambda_h |\Phi|^4 - \mu^2 |\Phi|^2 ,
\label{eq:portalpotential}
\end{align}
where  $\lambda_{\rm \phi h}$ is the portal coupling and $\Phi$ is the Standard Model Higgs, which after electroweak symmetry breaking can be replaced with $\Phi \rightarrow (v_h + h)/\sqrt{2}$, where $h$ is the neutral component of the SM Higgs doublet and $v_h \simeq 246$ GeV.

In Appendix \ref{app:higgsportal} we give a complete treatment of Higgs-inflaton mixing, and point out that the Higgs-inflaton portal term does not introduce a substantial tree-level inflaton mass term in the parameter space under consideration. The Higgs boson's observed branching fractions already indicate with $2 \sigma$ certainty that it decays at least four-fifths of the time like a Standard Model Higgs boson. Thus it is appropriate to refer to a mostly-Higgs-like, and a mostly-inflaton-like mass eigenstate, since the mixing between the two must be small to fit observations. Consistent with Section \ref{sec:simplequarticmodel}, we designate the mass eigenstate which is mostly-inflaton as ``$m_\phi$," and the mass eigenstate which is mostly-Higgs-like as ``$m_h$." 

For states which are mostly Higgs and mostly inflaton, the mixing angle between the Higgs and inflaton gauge eigenstates is defined as
\begin{align}
{\rm tan}~(2\theta_\phi) \equiv \frac{2 \lambda_{\phi h} v_{h} v_{\phi}}{|m_h^2 - m_{\phi}^2|},
\label{eq:mixingsimp}
\end{align}
where we set $m_h \simeq 125.7$ GeV in calculations.\footnote{The Higgs-curvaton mixing angle ($\theta_\sigma$) is identically defined, with the replacement $\phi \rightarrow \sigma$, $i.e.$ $v_\phi \rightarrow v_\sigma$, $m_\phi \rightarrow m_\sigma$, and $\lambda_{\phi h} \rightarrow \lambda_{\sigma h}$.} Contributions to $m_\phi$ and $v_\phi$ from the Higgs portal interaction are negligible (see Appendix \ref{app:higgsportal}), and so the mass and vacuum expectation value of the mostly-inflaton state, $v_\phi$ and $m_\phi$, are given by Eqs.~\eqref{eq:vphi} and \eqref{eq:mphi}. The preceding definition of $\theta_\phi$ has been chosen, so that in the limit of small $\theta_\phi$, the mostly-inflaton state mixes less with the Higgs boson, whether $m_\phi > m_h$ or $m_\phi < m_h$. In other words, as $\theta_\phi \rightarrow 0$, the inflaton's decay width to Standard Model particles vanishes, regardless of whether the inflaton-like state is heavier or lighter than the Higgs-like state. 

\begin{figure}[t!]
\center
\includegraphics[scale=.65]{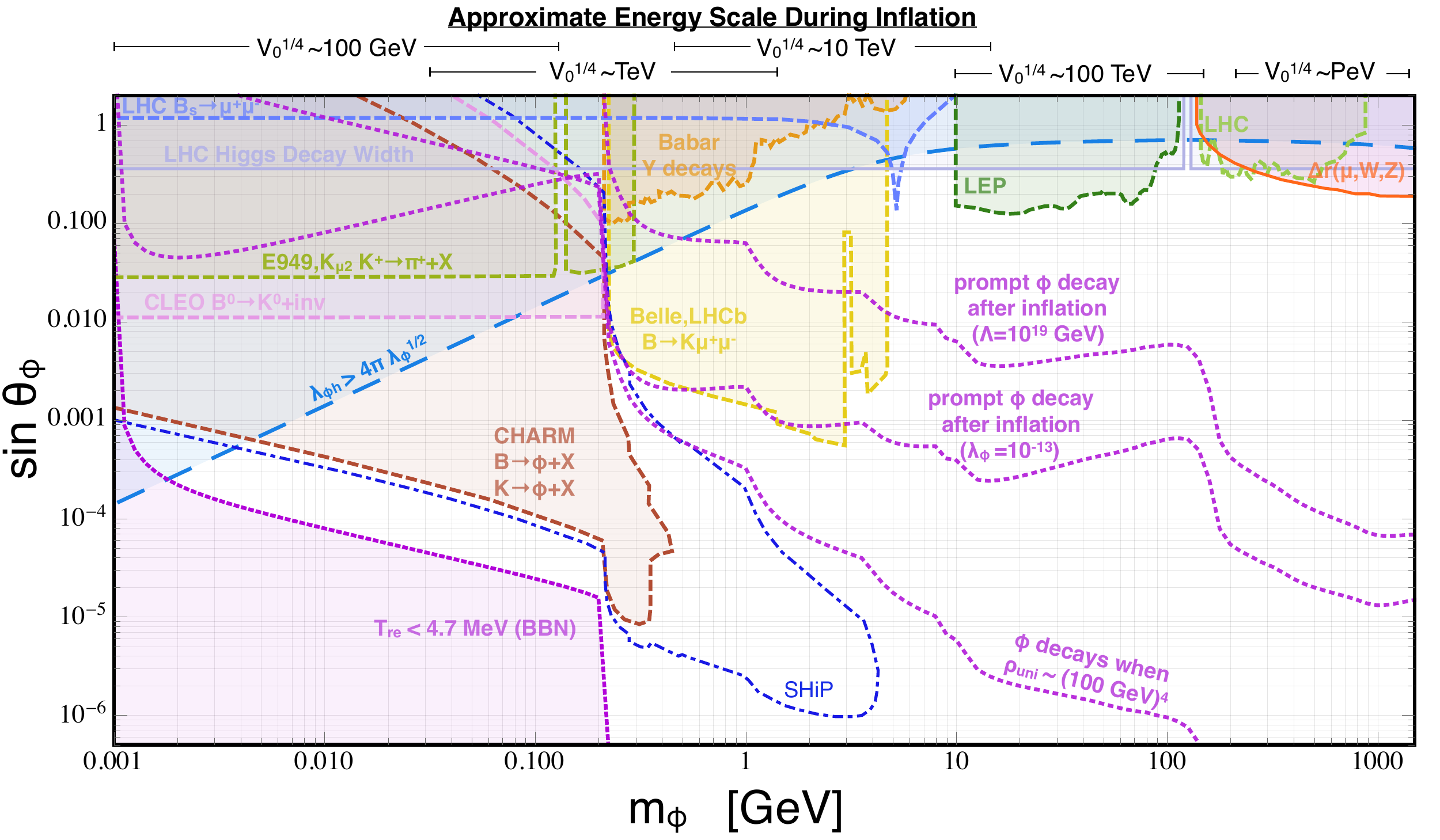}
\caption{Parameter space for low scale inflation, which reheats the universe through a Higgs portal coupling. Constraints from meson decay and collider searches are indicated with thick dashed lines. Indirect constraints from the muon's lifetime along with $W,Z$-boson masses ($\Delta r$) is indicated with a thin orange line, and the indirect constraint from the Higgs boson's decay width measured at the LHC is indicated with a thin gray line. The long-dashed blue line excludes parameter space where the Higgs-inflaton coupling ($\lambda_{\rm \phi h}$) spoils the flatness of the inflaton's potential during inflation. The dotted pink lines show parameter space where $\phi$ decays promptly at the end of inflation for $\Lambda = 10^{19}$ GeV and $\lambda_\phi=10^{-13}$, where $\phi$ decays when the energy density of the universe is $\sim (100 ~{\rm GeV)^4}$, and excludes where $\phi$ decays after before big bang nucleosynthesis. On top of the plot, the correspondence between the energy scale during inflation and the quartic inflaton mass is indicated. The range of inflationary energy scales is derived from relations shown in Figure \ref{fig:V0Lambda}; these ranges hold for a generic quartic inflaton, irrespective of the possible addition of a curvaton. With a curvaton model specified, the scale of inflation is more tightly predicted, see Table \ref{tab:paramaps}.}
\label{fig:portal}
\end{figure}

Examining the relative sizes of $v_h$, $m_h$, $m_\phi$ and $v_\phi$, for the parameter space shown in Figure \ref{fig:V0Lambda}, it is clear from Eq.~\eqref{eq:mixingsimp}, that because  $v_\phi \sim 10^3-10^9$ GeV, it is possible for $\theta_\phi$ to be sizable even if $\lambda_{\phi h}$ is small enough that it does not substantially correct the inflaton's quartic self-coupling ($\lambda_\phi$). The correction to $\lambda_\phi$ from the inflaton's portal coupling to the Higgs is
\begin{align}
\delta \lambda_{\phi} \sim \frac{\lambda_{\phi h}^2}{16 \pi^2},
\end{align}
up to $\mathcal{O}(10)$ logarithmic corrections. Therefore, to prevent the Higgs portal coupling from upsetting the flatness of the inflaton's potential, we can require $\lambda_{\phi h} < 4 \pi \sqrt{\lambda_\phi}$. In Figure \ref{fig:portal}, the resulting constraint on the size of the Higgs-inflaton mixing angle $\theta_\phi$ is shown in terms of $m_\phi$, with a long-dashed blue line.  It is interesting that, plotted in the (${\rm sin}~ \theta_\phi, m_\phi$) plane, the line $\lambda_{\phi h} = 4 \pi \sqrt{\lambda_\phi}$ is independent of the size of the quartic self-coupling, $\lambda_\phi$. This is because making the replacement $\lambda_{\phi h} \rightarrow 4 \pi \sqrt{\lambda_\phi}$ in the Higgs portal mixing angle, results in a mixing angle proportional to $m_\phi$, ${\rm tan ~(2\theta_\phi)} \propto \sqrt{\lambda_\phi} v_\phi \sim m_\phi$.

\subsection{Portal decay widths}
\label{subsec:portaldecays}
Assuming that the inflaton's only non-gravitational coupling to other particles is through its Higgs portal interaction, the decay widths of the mostly-Higgs and mostly-inflaton states are given by
\begin{align}
\Gamma_{h} & \simeq \Gamma_{\rm h,SM}(m_h) ~{\rm cos^2}~\theta_\phi \nonumber \\
\Gamma_{\phi} & \simeq \Gamma_{\rm h,SM}(m_\phi) ~{\rm sin^2}~\theta_\phi,
\label{eq:phidecays}
\end{align}
where $\Gamma_{\rm h,SM}(m)$ is the decay width for a boson of mass $m$, with Yukawa and gauge couplings identical to those of the Standard Model Higgs boson. (As in the prior subsection, all this discussion applies equally to the Higgs portal curvaton, with the replacement $\phi\rightarrow \sigma$ in all equations.) With this prescription, $\theta_\phi$ fully determines how fast the inflaton decays after inflation, and also how diminished the total decay width of the Higgs-like state will be, compared to Standard Model expectations. Because we are interested in parameter space where $\lambda_{\phi h} \ll 10^{-6}$, $\phi \rightarrow h h$ and $h \rightarrow \phi \phi$ decays are neglected. 

Many calculations of the partial decay widths of a Standard Model Higgs boson have been undertaken. Here we split the calculation of $\Gamma_{\rm h,SM}(m)$ into two pieces. For $m > 8 ~{\rm GeV}$, a scalar which couples like the Higgs boson, will decay predominantly to pairs of bottom quarks (and top quarks for $m > 350 ~{\rm GeV}$), and pairs of weak bosons. The decay of a heavy Standard Model Higgs boson has been calculated in a number of publications, including QCD corrections to hadronic decays of the Higgs, $e.g.$ Ref.~\cite{Djouadi:1995gt}. To compute $\Gamma_{\rm h,SM}(m)$ for $m > 8 ~{\rm GeV}$, we utilize output from \textsc{HDECAY} \cite{Djouadi:1997yw}, based on the calculations in \cite{Djouadi:1995gt}.

\begin{figure}[t!]
\center
\includegraphics[scale=.65]{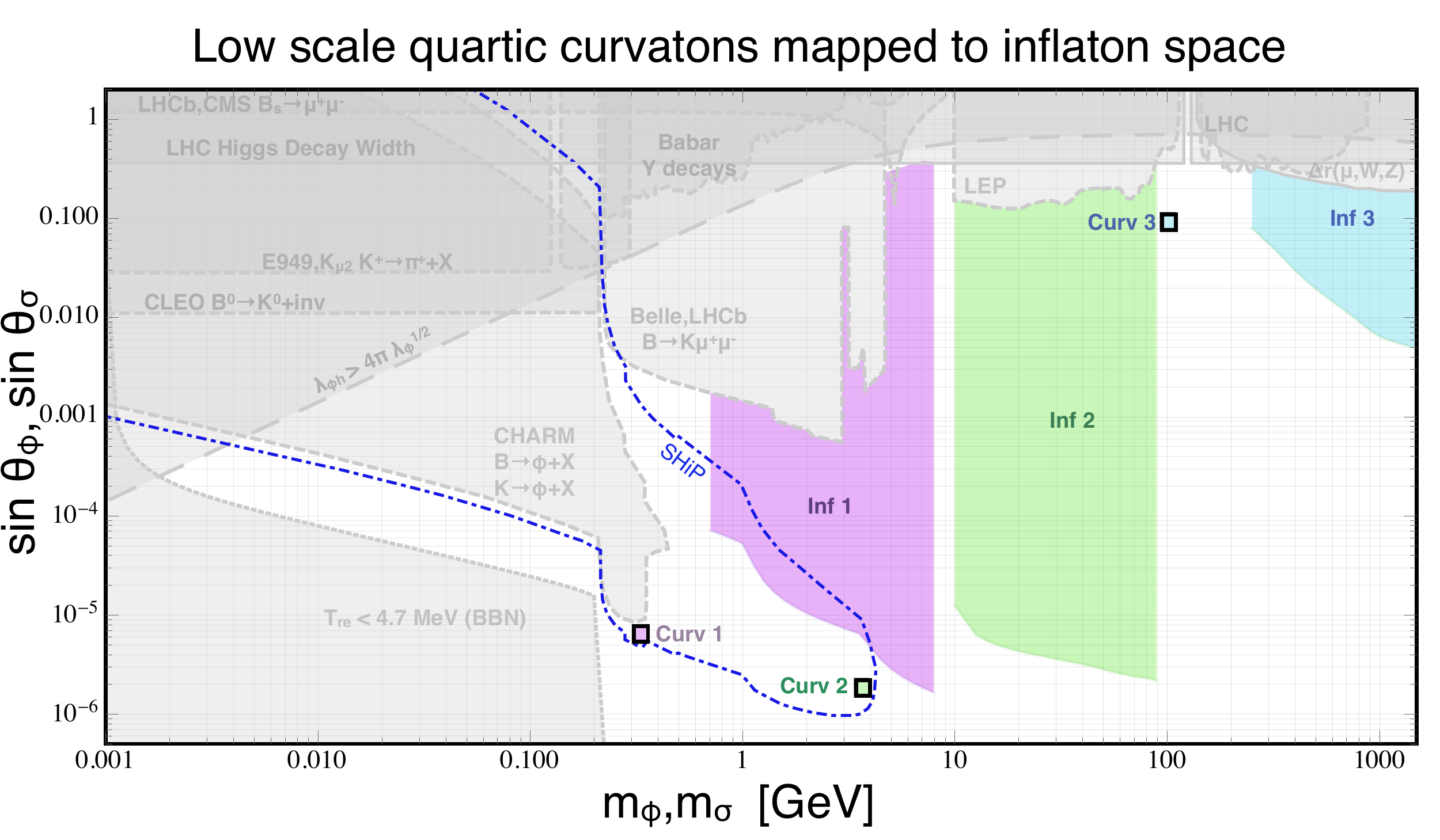}
\caption{The bounds and prospects are the same as in Figure \ref{fig:portal}, but here we show example quartic curvaton parameter points (Curv 1, Curv 2, Curv 3), alongside the corresponding predicted quartic inflaton parameter space (Inf 1, Inf 2, Inf 3), where these have been found using results in Section \ref{sec:maps}. Note that the curvaton and inflaton parameters roughly match those shown in Table \ref{tab:paramaps}.}
\label{fig:portalcti}
\end{figure}

For $m < 8 ~{\rm GeV}$, the Higgs-like scalar can decay to photons, leptons, and hadronic states, depending on whether each decay is kinematically permitted. The partial width for Higgs decay to photons is given by \cite{Ellis:1975ap,Djouadi:2005gj},
\begin{align}
\Gamma_{\rm h\rightarrow \gamma \gamma}(m) = \frac{\alpha_{\rm EM}^2m^3}{256 \pi^3 v_h^2} \left|\sum_f N_{\rm c} Q_{\rm f}^2 A_{1/2}(\tau_f)+ A_1(\tau_W) + \frac{m_W^2}{m_h^2}A_0(\tau_h)\right|^2,
\label{eq:photdecays}
\end{align}
where $\alpha_{\rm EM}$ is the fine structure constant, the displayed sum is over Standard Model fermions, $N_{\rm c}$ counts the colors of each fermion, $Q_{\rm f}$ is the electromagnetic charge of each fermion, $\tau_i \equiv m^2/4m_i^2$ where $m_i$ is the mass of particle $i$, and the loop amplitude functions,
\begin{align}
A_{1/2} (\tau)& = 2\tau^{-2}\left[\tau +(\tau -1)f(\tau)\right], \nonumber \\ A_{1} (\tau)& = -\tau^{-2}\left[2\tau^2+3\tau+(6\tau -3)f(\tau)\right] \\ A_{0} (\tau)& = -\tau^{-2}\left[\tau-f(\tau)\right], \nonumber
\end{align} 
with the scaling function $f(\tau)$ given by
\begin{align}
f(\tau) = \begin{cases}
               {\rm arcsin}^2 ~(\sqrt{\tau}),&~~~~~~~\tau \leq 1\\
              -\frac{1}{4}\left({\rm log} ~\left[\frac{1+\sqrt{1-\tau^{-1}}}{1-\sqrt{1-\tau^{-1}}}\right] -i\pi\right)^2,&~~~~~~~\tau > 1.
            \end{cases}
\end{align}
Following \cite{Dolan:2014ska}, in the preceding expressions we use the pion mass and kaon mass for the up, down, and strange quarks, $i.e.$ $\tau_{\rm u} = \tau_{\rm d}= m^2/4m_\pi^2$, $\tau_{\rm s} = m^2/4m_K^2$. This mass choice results in decay widths that match results from chiral perturbation theory \cite{Leutwyler:1989tn,Donoghue:1990xh}. 

The decay width to Standard Model leptons is given by
\begin{align}
\Gamma_{\rm h\rightarrow \ell \ell}(m) = \frac{\alpha_{\rm EM}^2m_\ell^2 m}{8 \pi v_h^2} \beta_\ell,
\end{align}
where $\beta_i \equiv \left(1-\tau_i^{-1} \right)^{3/2} \Theta \left(m-2m_i \right)$, with the Heaviside Theta function accounting for decays that are kinematically forbidden.

To compute the hadronic decays of a light Higgs-like scalar, we follow the treatment of \cite{McKeen:2008gd}, which matches the perturbative spectator model for Higgs decays \cite{Gunion:1989we}, onto chiral perturbation theory evaluated at the QCD scale. With this prescription, the relative leptonic and hadronic decay widths are given by
\begin{align}
\Gamma_{\rm h\rightarrow e e}   &:\Gamma_{\rm h\rightarrow \mu \mu}   : \Gamma_{\rm h\rightarrow \tau \tau} : \Gamma_{\rm h\rightarrow g g} : \Gamma_{\rm h\rightarrow \pi \pi} : \Gamma_{\rm h\rightarrow KK} : \Gamma_{\rm h\rightarrow \eta \eta} : \Gamma_{\rm h\rightarrow DD} \nonumber \\ &= m_e^2 \beta_e :m_\mu^2 \beta_\mu :m_\tau^2 \beta_\tau : \left( \frac{\alpha_s m}{3 \pi}\right)^2 \left(6-2\beta_\pi-\beta_K\right): \nonumber \\ &3(m_u^2+m_d^2)\beta_\pi : \frac{27}{13} m_s^2 \beta_K : \frac{12}{13} m_s^2 \beta_\eta: m_c^2 \beta_D :m_b^2 \beta_B,
\label{eq:hadecays}
\end{align}
where $e,\mu,\tau,\pi,K,\eta,D,B,g,u,d,s,c$ indicate the electron, muon, tau, pi-meson, k-meson, eta-meson, D-meson, B-meson, up-quark, down-quark, strange-quark, and charm-quark of the Standard Model. Matching to chiral perturbation theory at the QCD scale, for the hadronic decay calculation we take $m_u=m_d = 50 ~{\rm MeV}$, $m_s = 450 ~{\rm MeV}$, $\alpha_s =0.47$, and the current quark and meson masses given in \cite{Agashe:2014kda}.

\subsection{Finding low scale inflation through a Higgs portal}

\begin{figure}[t!]
\center
\includegraphics[scale=.65]{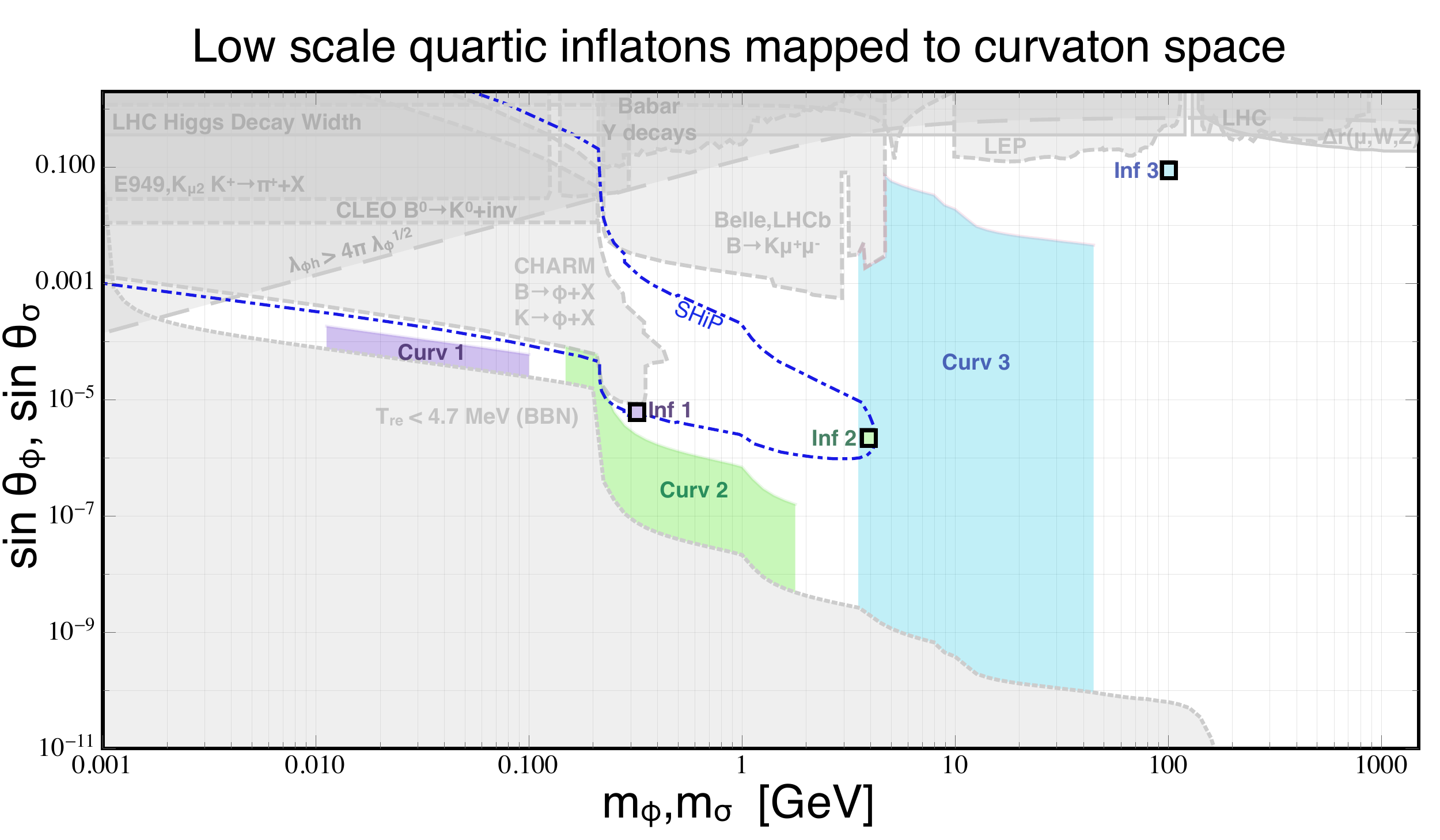}
\caption{The bounds and prospects are the same as in Figure \ref{fig:portal}, but here we show example quartic inflaton parameter points (Inf 1, Inf 2, Inf 3), alongside the corresponding predicted quartic curvaton parameter space (Curv 1, Curv 2, Curv 3), where these have been found using results in Section \ref{sec:maps}. Note that the inflaton and curvaton parameters roughly match those shown in Table \ref{tab:paramaps}.}
\label{fig:portalitc}
\end{figure}

Using Eqs.~\eqref{eq:phidecays}--\eqref{eq:hadecays} to calculate the decay rate of $\phi$, cosmological limits can be placed on Higgs portal parameter space, for models of low scale inflation that reheat by coupling to the Higgs boson. First, there is an absolute lower bound on the inflaton's (or curvaton's) decay rate, from the requirement that the universe reheat before big bang nucleosynthesis, namely that decay occurs before $ T_{\rm BBN} \simeq 4.7~{\rm MeV}$, which excludes parameter space in the lower left of Figure \ref{fig:portal}. Similarly, one might require that the inflaton decay before the universe reaches a density of $\rho_{\rm uni} \sim {\rm (100~ GeV)^4}$, which is necessary for some cosmologies that incorporate electroweak baryogenesis. Using the relation $\Gamma_\phi \sim H = \sqrt{\rho_{\rm uni}/3 M_{\rm p}^2}$, Figure \ref{fig:portal} shows parameter space consistent with $\phi$ decay before $\rho_{uni} \sim {\rm (100~GeV)^4}$ with a pink dotted line. Next, one might require that $\phi$ decay promptly at the end of inflation, $i.e.$ $\Gamma_\phi \sim \sqrt{V_0/3 M_{\rm p}^2}$. For a given value of $m_\phi$, specifying either $\lambda_\phi \sim 10^{-13}$ or $\Lambda \sim 10^{19} ~{\rm GeV}$ uniquely determines $V_0$, using Eqs.~\eqref{eq:simplam} and \eqref{eq:mphi}. In Figure \ref{fig:portal}, we show parameter space consistent with nearly instantaneous reheating after inflation for $\lambda_\phi \sim 10^{-13}$ and $\Lambda \sim 10^{19} ~{\rm GeV}$. Altogether, most low scale quartic inflaton models that reheat through the Higgs portal could be probed by more extensive Higgs measurements and searches for low-mass scalars.

Next, relations derived in Section \ref{sec:maps} have shown a characteristic mass and decay width spectrum for a quartic inflaton-curvaton pair. These inflaton-curvaton pairs could become apparent through Higgs portal interactions. Figures \ref{fig:portalcti} and \ref{fig:portalitc} each indicate three points in Higgs portal parameter space, identify them as a quartic curvaton or inflaton, respectively, and show where a corresponding low scale quartic inflaton or curvaton would appear. The curvaton and inflaton points shown in Figures \ref{fig:portalcti} and \ref{fig:portalitc} match parameters given in Table \ref{tab:paramaps}. Here the decay width bounds have been recast as bounds on $\theta_{\phi},\theta_{\sigma}$, by using the definition of the mixing angle and the portal decay width, Eqs.\eqref{eq:mixingsimp} and \eqref{eq:phidecays}, and the decay width of a Higgs-like scalar for a given mass, detailed in \ref{subsec:portaldecays}.

Looking at Figures \ref{fig:portalcti} and \ref{fig:portalitc}, it is apparent that in some cases, an extended run of the SHiP \cite{Alekhin:2015byh} experiment would suffice to uncover both a quartic inflaton and curvaton field. If a scalar state is discovered at SHiP with a mass $0.3$ -- $4$ GeV, then if the state is a quartic inflaton, one should expect a quartic curvaton in the mass range $0.01$ -- $1$ GeV. On the other hand a quartic inflaton should show up in the mass range $1$ -- $100$ GeV if the discovered $0.01$ -- $1$ GeV scalar is a quartic curvaton. Furthermore, it is apparent from Figure \ref{fig:portalitc}, which plots down to very small mixing angles ($\theta_\phi,\theta_\sigma \sim 10^{-11}$), that while substantially expanded meson production would be necessary, as is planned at experiments like SuperKEKB and SHiP \cite{Alekhin:2015byh,Akeroyd:2004mj,Browder:2007gg}, future Higgs portal searches could conceivably be sensitive to cosmological scalars with decay widths corresponding to BBN reheat temperatures, and thereby discover or rule out classes of low scale inflatons and curvatons.

\subsection{Experimental probes of MeV-TeV mass Higgs portal scalars}

Figures \ref{fig:portal}, \ref{fig:portalcti}, and \ref{fig:portalitc} display bounds on Higgs portal parameter space from direct and indirect measurements of the Standard Model Higgs boson along with searches for new scalars at meson factories. This section details experimental probes of Higgs portal scalars, beginning with searches for higher mass states.

The addition of a Higgs portal singlet scalar with a large mass can alter the relationship between the $W$-boson mass, the $Z$-boson mass, the Fermi constant, and the decay rate of the muon. If the Higgs portal singlet is massive enough, its corrections to the electroweak bosons' self-energy are too large, given the observed lifetime of the muon, leading to a bound ${\rm sin} ~\theta_\phi \lesssim 0.2$ for $m_\phi \gtrsim 300$ GeV at $95 \%$ confidence \cite{Lopez-Val:2014jva}. This indirect bound is shown in Figure \ref{fig:portal} with a thin orange line.

A $m_\phi \gtrsim 140$ GeV Higgs portal scalar can be detected at the LHC, largely through its decays to leptons, $pp \to \phi \to ZZ \to 4 \ell$ and $pp \to \phi \to WW \to \ell \nu \ell \nu$ \cite{Khachatryan:2015cwa,Aad:2015kna}. Statistical combinations of ATLAS and CMS results \cite{Robens:2015gla} yields the tightest bound on a Higgs portal scalar in the mass range $m_\phi \sim 150-250$ GeV (shown in dashed green in Figure \ref{fig:portal}). 

A portal interaction would diminish the effective width of the Higgs boson that has been observed at the LHC. ATLAS and CMS have placed the most restrictive lower bound on the Higgs width using Higgs decays to leptons and photons, $(h \rightarrow ZZ \rightarrow 4 \ell)$ and $(h \rightarrow \gamma \gamma)$ \cite{Khachatryan:2016vau}. The combined limit on the signal strength of the Higgs ($\mu_{\rm higgs} \equiv \sigma_{\rm meas.} / \sigma_{\rm SM}$) is $\mu_{\rm higgs} > 0.87,~{\rm at}~95\%~{\rm confidence}$, which corresponds to an upper bound of ${\rm sin}~\theta_\phi < 0.36$ at 95\% confidence. This indirect bound would not be sensitive to a Higgs portal scalar mass-degenerate with the observed Higgs boson, and applies to $m_\phi < 120~{\rm GeV}$ and $m_\phi >130~{\rm GeV}$. 

A combination of searches at LEP \cite{Schael:2006cr} did not observe a light Higgs-like state in ($e^+ e^- \rightarrow  Z h \rightarrow Z b \bar{b}$) and ($e^+ e^- \rightarrow Z h  \rightarrow  Z\tau^+ \tau^-$) over a mass range $\sim 10-100$ GeV. Recasting the 95\% confidence bound set by LEP limits Higgs portal couplings to ${\rm sin} ~\theta_\phi \lesssim 0.2$ for $m_\phi \sim 10-100$ GeV.

Because a Higgs portal scalar couples to quarks, it contributes to the amplitude for Standard Model meson decay. The portal scalar considered here couples to Standard Model fermions with the same proportions as a Standard Model Higgs boson with mass $m_\phi$, as described in Section \ref{subsec:portaldecays}. Therefore, the Higgs portal scalar preserves quark flavor at tree level, but can induce meson decay processes like $B \rightarrow K \mu^+ \mu^-$ at loop-level, through ``penguin" diagrams containing internal $W,Z$ boson lines. References~\cite{McKeen:2008gd,Clarke:2013aya,Dolan:2014ska,Alekhin:2015byh} have catalogued the bounds on Higgs portal scalars from loop-induced decays of mesons in the mass range $m_\phi \sim 0.001-5$ GeV, displayed in Figures \ref{fig:portal}--\ref{fig:portalitc}.

\section{Conclusions}
\label{sec:conclusions}

The spectrum of tensor perturbations produced by low scale inflation models -- and by extension the energy scale of inflation -- is too small to be uncovered by cosmological surveys. However, this study has shown that low scale inflatons which roll to large field values, and to some approximation, the corresponding energy scale during inflation, can be probed at colliders and meson factories. 

Broadly speaking, low scale inflation deserves attention, because recent cosmological surveys have begun ruling out high scale models. In addition, low scale inflation may be a necessity if our universe contains axions. From a theoretical standpoint, low scale inflation can be described with a low-energy effective field theory, whereas high scale inflation requires suppression of radiative corrections to trans-Planckian dynamics. 

The possibility of finding an inflaton at a collider may seem exotic, partly owing to an assumption that inflatons are too heavy for terrestrial production. In the regime of large field inflation, this is often true (in the case of $m^2 \phi^2$ large field inflation, $m \sim 10^{-6}$ $M_{\rm p}$). However, in the case of low scale, small field inflation, it is natural to suppose that the inflaton begins with a nearly null field value, subsequent to a phase transition.  In this ``hilltop" case, the inflaton rolls down its potential, settling at a large vacuum expectation value.  This large VEV, along with the tiny self-couplings required of a low scale slow-roll inflaton, result in a small inflaton mass detectable at a low energy experiment. This study has shown that a small field quartic hilltop potential implies an inflaton mass ranging from an MeV to PeV, corresponding to an inflationary energy scale ranging from GeV to EeV, which can be probed at terrestrial collider experiments through a Higgs portal interaction.

The Higgs portal cosmology and low-energy phenomenology developed for a simplified quartic hilltop model of inflation, could be applied to broader classes of small field inflation that initiate with a nearly null inflaton field value, and roll to a large vacuum expectation value. It is particularly interesting that, owing to its large vacuum expectation value at the end of inflation, such an inflaton can have a tiny coupling to the SM Higgs boson ($\lambda_{\phi h } \ll 10^{-6}$), yet still have a sizable enough mixing to rapidly reheat the universe, all without spoiling the flatness of the inflaton's potential through radiative corrections. This constitutes one clear mechanism for a low scale inflaton with an extremely flat potential to substantially couple to the Standard Model, without fine-tuning. This also reinforces the cosmological import of Higgs portal scalar searches, both at high energy colliders like the LHC, and in flavor-violating meson decays at experiments like KEKB, BEPC, and SHiP.

Intriguingly, this study has demonstrated that once a complete cosmology is specified, and primordial perturbations accounted for, it is possible to make sensible predictions for the relative masses and decay widths of scalars associated with low scale inflation. The fairly simple case of a quartic hilltop inflaton paired with a quartic curvaton has been studied in detail, and maps between the masses and decay widths of each have been charted. The same methods can be used to infer the energy density during low scale inflation. Specifically, using a simplified quartic hilltop inflaton in Section \ref{sec:simplequarticmodel}, the requirement that the inflaton's potential be stabilized by operators in an effective field theory with a sub-Planckian cutoff, is sufficient to map the mass of the inflaton to the energy scale during inflation, to within roughly an order of magnitude.  After adding a realistic curvaton cosmology, twinned with the requirement that the average equation of state and temperature during reheating have physically permissible values detailed in sections \ref{sec:reheatingconstraints} and \ref{sec:curvaton}, this map tightened -- as shown by the restricted values for the inflationary energy density and number of efolds given in Table \ref{tab:paramaps}.  While this study has focused on a low scale quartic hilltop inflaton, the same cosmological analysis could be applied to any low scale inflaton (or curvaton) model, which maintains the necessary flatness of its potential with an initially small field value and small self-couplings.  When these scalars roll to their minima and acquire large vacuum expectation values, the same reheating and perturbation considerations which constrained the mass and decay width of quartic hilltop inflaton and curvatons, apply to other low scale inflatons and curvatons.  

Particularly, it will be interesting to extend these techniques to additional hilltop, pseudo Nambu Goldstone boson, and inflection point models of low scale inflation, to further determine how meson factories, high energy colliders, and other experimental probes of scalar fields could unmask low scale inflation.
\\
\\
{\bf Acknowledgments}
We thank Rouzbeh Allahverdi, Daniele Alves, Kohei Kamada, David McKeen, Tilman Plehn, Jessie Shelton, and Nirmal Raj for useful discussions. JB thanks Los Alamos National Laboratories (LANL), the Center for Theoretical Underground Physics (CETUP), Heidelberg University, and the Aspen Center for Physics, which is supported by National Science Foundation grant PHY-1066293, for hospitality while portions of this work were completed. This work was partially supported by the National Science Foundation under Grants No. PHY-1417118 and No. PHY-1520966.

\appendix

\section{Small field hilltop models} \label{app:othermodels}

In this section we examine a number of small field hilltop models, quantifying fine-tuning in quadratic, cubic and quartic hilltop inflation. Partly, this will justify the choice of a quartic inflaton plus quartic curvaton model, as the simplest practicable case of small field inflation that is driven by a single Lagrangian term.\footnote{It will be interesting to extend results in this paper to ``inflection point" models \cite{Allahverdi:2006we,Baumann:2007np,Allahverdi:2011su}, where a few Lagrangian terms driving inflation are tuned against each other to produce the observed spectrum of primordial perturbations.} 

Small field inflation requires an especially flat potential (compared to large field inflation), so that it is natural to consider a hilltop potential, $e.g.$ of the form $V(\phi) = - \lambda_n \phi^n + V_0$. For this potential, a very flat portion exists at the origin of field space. Typically, the self-coupling terms of a hilltop inflaton must be very small to permit inflation. To understand why, it is instructive to consider a scalar potential familiar to particle theorists, the potential of the Higgs boson in the Standard Model, and examine why, with its comparatively large self-coupling, the Higgs potential does not permit hilltop inflation. (Sometimes the Higgs boson, with an additional large coupling to gravity, is considered as the inflaton \cite{Bezrukov:2007ep}. In that non-minimally coupled case, the Higgs begins inflation at very large field values. Here we study the Higgs hilltop inflation scenario, where the Higgs has no new coupling to gravity, and has a nearly null initial field value.)

To attempt Higgs hilltop inflation, one considers a Higgs rolling from its hilltop at a nearly null value $h \gtrsim 0$, to its electroweak minimum $h \sim 246~{\rm GeV}$. First we must address how the Higgs might have a nearly null initial field value. One might suppose that after electroweak symmetry breaking, the Higgs automatically starts near the top of hill. However, assuming a Standard Model-like phase transition, the thermal fluctuations of the Higgs would be too large ($\mathcal{O}(100~{\rm GeV})$) and inflation would not occur. For the moment we will ignore thermal fluctuations, and assume that the Higgs field can begin with an arbitrarily uniform null field value; some discussion about how this can be achieved for hilltop potentials was provided in Section \ref{sec:simplequarticmodel}. However, even setting aside thermal fluctuations, another restriction on a nearly null initial field value, comes from fluctuations in scalar fields induced by the de Sitter (inflationary) space they presumably occupy. In other words, if we specify that the initial Higgs field value is very nearly null, we may violate the intrinsic quantum uncertainty of a scalar field in de Sitter space. A scalar field in a de Sitter space with Hubble constant $H$ fluctuates as $\delta h \sim H / 2 \pi$. We will see that the initial field value necessary for $20-50$ efolds of Higgs hilltop inflation is much smaller than this, $h_{ini}^{(\rm 60~efolds)} \ll H/2 \pi$.

We begin with a toy Higgs hilltop potential,\footnote{$N.b.$, this treatment is for illustrative purposes, and is completely independent from the Higgs portal considerations in Section \ref{sec:inflatonhiggsportal} and Appendix \ref{app:higgsportal}.}
\begin{align}\label{eq1}
V = V_0 - \mu_h^2 h^2 + \lambda h^4,
\end{align}
with $\mu_h \simeq v \sqrt{\lambda}$ where $v$ is the Higgs vev and $V_0 = \frac{v^4 \lambda}{4}$ such that when the Higgs is sitting at its electroweak minimum, it does not over-contribute to the dark energy of the universe ($V(h_{min}) \simeq 0$). We can compute how close to $h = 0$ the Higgs field must be in order for the universe to inflate by 60 efolds. The number of efolds is given by
\begin{align}
N = \frac{1}{M_{\rm p}^2} \int^{h_{60}}_{h_{end}} \frac{V}{V_{h}} d h \simeq \frac{v^2}{4 M_{\rm p}^2} {\rm log} \frac{h_{end}}{h_{60}},
\end{align}
where we have dropped the quartic Higgs term, which will be irrelevant at small field values, and defined the Higgs field value at the end of inflation $h_{end}$, and at the start of 60 efolds of inflation, $h_{60}$. We determine $h_{end}$ by solving for the field value at which $\epsilon=1$, and use $v = 246$ GeV to obtain
\begin{align}
h_{60} \simeq 10^{-17} \, e^{- 10^{34} } ~{\rm GeV},
\end{align}
which is absurdly infinitesimal compared to quantum fluctuations in the Higgs field,
\begin{align}
\frac{H}{2 \pi} \approx \frac{\nu^2 \sqrt{\lambda}}{ 4 \pi \sqrt{3} M_{\rm p}} \simeq 4.1 \times 10^{-17} ~{\rm GeV}.
\end{align}
Therefore to inflate our universe to the extent implied by CMB observations, the initial Higgs field value would need to be specified well within the de Sitter quantum uncertainty limit. Conversely one might ask the maximum number of efolds achievable with the Higgs hilltop potential, while staying within the de Sitter quantum uncertainty limit. The answer is tiny, the maximum number achievable is $N_{max}^{(\rm Higgs ~hilltop)} = 10^{-32}$. This means that the Standard Model Higgs potential would not generate enough inflation in a hilltop scenario, as a consequence of its relatively large self-coupling.

We now discuss fine-tuning and primordial perturbations generated by small field hilltop models of inflation, where the involved scalar fields have tiny self-couplings.

\subsection*{$\phi^2$ hilltop inflation}

If practicable, it would be preferable to consider a hilltop potential where a $-\phi^2$ term drives inflation, $e.g.$
\begin{align}
V = V_0 - \frac{1}{2} m^2 \phi^2 + \frac{1}{\Lambda} \phi^6,
\end{align} 
similar to the $\phi^4$ case utilized in the bulk of the paper. The advantage of such a potential is that, in the absence of explicit quartic and cubic terms, this potential is technically natural. However, with the additional requirement that inflation ceases when the inflaton rolls to the minimum of this potential, under the stipulation that $\Lambda \leq 10^{19}$ GeV, $m$ becomes too large to be compatible with inflation. Specifically, one finds that the power spectrum resulting from such a potential is too large, and that generating 60 efolds of inflation typically requires $\phi_* < \frac{H_*}{2 \pi}$, in violation of the de Sitter space quantum uncertainty limit discussed above. 

This model can be mended by suppressing the high scale operator,
\begin{align}
V = V_0 - \frac{1}{2} m^2 \phi^2 + \frac{\delta}{\Lambda^2} \phi^6
\end{align} 
with $\delta \ll 1$. However, this implies that corrections from trans-Planckian dynamics are somehow suppressed.

The other option is to add in a negative $\phi^4$ term,
\begin{align}
V = V_0 - \frac{1}{2} m^2 \phi^2 - \frac{1}{4} \lambda_\phi \phi^4 + \frac{1}{\Lambda^2} \phi^6,
\end{align} 
The added $ \lambda_\phi \phi^4/4$ term generates a contribution to the mass at loop order, and because (unsurprisingly), its value is equal or greater than that given in \eqref{eq:simplam}, fine-tuning of the model is not ameliorated (as compared to just using the quartic term as the dominant term during inflation). 

\subsection*{$\phi^4$ hilltop inflation}

To quantify fine-tuning in the quartic case, it is required that the mass term for the inflaton while rolling through its pivot scale is no more then $10$ \% of the quartic term, $m_{\phi,*}^2 < m_{\phi, \rm max}^2 \equiv 0.1 \lambda_\phi \phi_{*}^2/4$, where we remind the reader that $m_{\phi,*}$ is the sum of bare and loop contributions to the inflaton's mass. Note if the preceding inequality is satisfied at the pivot scale, then it is automatically satisfied at larger field values, during and after inflation ($\phi$ grows during and after inflation). We then compare this to the mass generated at one-loop order,
\begin{align}
m^2_{\phi, \rm loop} \sim \frac{\lambda_\phi \Lambda^2}{16 \pi^2} ~ .
\end{align}
Using Eqs.~\eqref{eq:simplam} and \eqref{eq:Nefforsimpmodel}, we define the ratio of $m_{\phi , \rm max}$ and $m_{\phi, \rm loop}$ as the level of fine-tuning in the theory,
\begin{align}
\frac{m_{\phi ,\rm max}}{m_{\phi, \rm loop}} \simeq 0.72 \frac{V_{0 \phi}^{\frac{1}{3}}}{M_{\rm p} \Lambda^{\frac{1}{3}} \sqrt{N_*}}
\end{align}
Without specifying some theory that forbids a quadratic term in the Lagrangian, fine-tuning considerations favor a scale of inflation, $V_{0 \phi}^{1/4}$, close to the cutoff scale $\Lambda$, and favors larger scale inflation. For example, for $V_{0 \, \phi}^{\frac{1}{4}} = \Lambda = 10^{15} $ GeV and $N_* = 30$ (where the inflaton's mass at its minimum will be $m_\phi \sim 100$ GeV), we find $\frac{m_{\phi, \rm max}}{m_{\phi, \rm loop}} \sim 10^{-4}$.

\subsection*{$\phi^3$ hilltop}

It can be shown that fine-tuning is not greatly improved in the case of small field hilltop inflation driven by a $\phi^3$ term. One might expect fine-tuning decreases, because the cubic one-loop-induced mass term depends on two factors of the cubic coupling (instead of one in the case of the quartic). For the potential $V= V_0 - \frac{1}{3} g \phi^3 + \frac{\phi^5}{\Lambda} $,\footnote{Here the potential is stabilized when $\phi$ takes positive field values. An additional $\phi^6$ term might be included to stabilize the potential for negative $\phi$ field values.} the leading loop contribution to the mass is
\begin{align}
m_{\phi, \rm cubic~loop}^2 \simeq \frac{g^2 \Lambda^2}{9 \cdot 2^4 \pi^2 m_0^2},
\end{align}
where $g$ is the dimensionful coupling of the $\phi^3$ term. Comparing this to the maximum mass as defined in the prior section,
\begin{align}
m_{\phi, \rm max ~cubic }^2 \equiv \frac{1}{3} g^3 \phi_* \, .
\end{align}
Taking the ratio of $m_{\phi, \rm max ~cubic }$ and $m_{\phi, \rm cubic~loop}$, using that for hilltop $\phi^3$ inflation, $\phi_* \approx \frac{V_{0 }}{g M_{\rm p}^2 N_*}$,
\begin{align}
\frac{m_{\phi, \rm max ~cubic }}{m_{\phi, \rm cubic~loop}} \simeq 1.59 \sqrt{\frac{V_0}{g M_{\rm p}^2 \Lambda N_*}} \, .
\label{eq:cubictune}
\end{align}
The requirement that $V(\phi_{min})=0$ determines $g$ in terms of $V_0$ and $\Lambda$, as for the quartic self-coupling in Section \ref{sec:simplequarticmodel},
\begin{align}
g = \left(\frac{75 \sqrt{5}}{2} \frac{V_0}{\Lambda^{3/2}} \right)^{\frac{2}{5}} \, .
\end{align}
Inserting this into Eq.~\eqref{eq:cubictune}, 
\begin{align}
\frac{m_{\phi, \rm max ~cubic }}{m_{\phi, \rm cubic~loop}} \simeq 0.7 \frac{V_0^{3/10}}{\Lambda^{1/5} M_{\rm p} \sqrt{N_*}} \, .
\end{align}
For fixed $V_0, \Lambda$, the tuning of the cubic hilltop model does not improve over the quartic case. For example for $V_0^{1/4} = \Lambda = 10^{15}$GeV and $N_*=30$, the tuning is roughly the same as for the quartic potential, $\frac{m_{\phi, \rm max ~cubic}}{m_{\phi, \rm cubic~loop}} \sim 10^{-4}$.

\section{Higgs-inflaton and Higgs-curvaton portal particulars}
\label{app:higgsportal}
In what follows, as in Section \ref{sec:inflatonhiggsportal}, we address quartic inflaton-Higgs mixing, with the understanding that an identical treatment applies to quartic curvaton-Higgs mixing. For the potential of Eq.~\eqref{eq:portalpotential}, the vacuum expectation values of $h$ and $\phi$ are
\begin{align}
v_{\rm h}^2 = \frac{\mu^2 +\lambda_{\rm \phi h } v_{\rm \phi}^2}{2 \lambda_{\rm h}},~~~~~~v_{\phi}^2 = \frac{\Lambda^2 \lambda_{\rm \phi} \left(1+\sqrt{1-\frac{48 \lambda_{\rm \phi h} v_{\rm h}^2 }{\Lambda^4\lambda_\phi^2}} \right)}{2 \sqrt{3}},
\end{align}
the mass matrix for the neutral Higgs component and inflaton is given by
\begin{align}
\mathcal{M}^2 = 
\left(
\begin{array}{cc}
2\lambda_{\rm h} v_{\rm h}^2 & \lambda_{\rm  \phi h} v_{\rm h} v_{\rm \phi}  \\
\lambda_{\rm \phi h} v_{\rm h} v_{\rm \phi} & 3\lambda v_{\phi}^2+\frac{30 v_{\phi}^4}{\Lambda^2} +2 \lambda_{\rm \phi h} v_{\rm h}^2
\end{array} 
\right),
\end{align}
for which the mass eigenstates are
\begin{align}
M_{1,2}^2 = \frac{1}{2} \left( 2\lambda_{\rm h} v_{\rm h}^2 + 3\lambda_\phi v_{\phi}^2+\frac{30 v_{\phi}^4}{\Lambda^2}+2 \lambda_{\rm \phi h} v_{\rm h}^2 \mp \sqrt{4\lambda_{\rm \phi h} v_{\rm h} v_{\rm \phi} + \left(2\lambda_{\rm h} v_{\rm h}^2 - 3\lambda v_{\phi}^2-\frac{30 v_{\phi}^4}{\Lambda^2}-2 \lambda_{\rm \phi h} v_{\rm h}^2 \right)^2} \right).
\end{align}

A common definition for the mixing angle between the two Higgs portal mass eigenstates ($S_{1,2}$) is $\alpha$, such that
\begin{align}
S_1 &= h~ {\rm cos}~\alpha + \phi~ {\rm sin}~\alpha \nonumber \\
S_2 &= h~ {\rm sin}~\alpha + \phi~ {\rm cos}~\alpha ,
\end{align}
where in turn $\alpha$ is given as
\begin{align}
{\rm tan} (2 \alpha) = \frac{2\lambda_{\rm \phi h} v_{\rm h} v_{\rm \phi} }{3\lambda v_{\phi}^2+\frac{30 v_{\phi}^4}{\Lambda^2}+2 \lambda_{\rm \phi h} v_{\rm h}^2-2\lambda_{\rm h} v_{\rm h}^2}.
\end{align}

In this study, it is convenient to define the mixing angle differently, as discussed in text surrounding Eq.~\eqref{eq:mixingsimp}. Specifically, we wish to define the mixing angle so that as the mixing angle vanishes, so too does the decay width of the mostly-inflaton state to Standard Model particles. Because the mass of the Higgs boson has been measured, one of the mass eigenstates $M_{1},M_2$ must be $\simeq 125$ GeV. We consistently refer to the Higgs-like mass state as ``$m_h$" in this document, the inflaton-like state as $m_\phi$, and the mixing angle between the Higgs-like and inflaton-like states as $\theta_\phi$, where
\begin{align}
{\rm tan} (2 \theta_\phi) \equiv \frac{2\lambda_{\rm \phi h} v_{\rm h} v_{\rm \phi} }{|3\lambda v_{\phi}^2+\frac{30 v_{\phi}^4}{\Lambda^2}+2 \lambda_{\rm \phi h} v_{\rm h}^2-2\lambda_{\rm h} v_{\rm h}^2|} \simeq \frac{\lambda_{\rm \phi h} v_{\rm h} v_{\rm \phi} }{|m_\phi^2-m_h^2|},
\end{align}
where we have dropped the portal mass contribution, $2 \lambda_{\rm \phi h} v_{\rm h}^2$, in the final expression. This term will not contribute substantially to the inflaton's mass at its minimum for two reasons. The first reason, is that by necessity, the inflaton's effective mass term at the outset of inflation must be much smaller than its quartic term, $\lambda_\phi \phi_*^2 \gg m_{\phi,*}^2$, as detailed in Appendix \ref{app:othermodels}. One consequence is that the Higgs portal contribution to the inflaton mass must be much smaller than $m_{\phi,\rm max}$, which is smaller than $m_\phi$.  However, it should be stressed that the Higgs portal mass contribution ``$2 \lambda_{\rm \phi h} v_{\rm h}^2$" will be negligible anyway in most of the parameter space we consider, from the requirement $\lambda_{\phi h} \lesssim 10^{-6}$, discussed in Section \ref{sec:inflatonhiggsportal} (one might also consider whether $v_h \sim 0$ during inflation). The smallness of the Higgs portal operator, $\lambda_{\phi h} \lesssim 10^{-6}$, was required so that the Higgs portal quartic would not upset the inflaton's self-quartic coupling through radiative corrections. In fact, in all un-excluded $m_\phi \gtrsim 0.05$ GeV inflaton-curvaton parameter space in Figures \ref{fig:portalcti} and \ref{fig:portalitc}, the portal contribution to the inflaton mass can be neglected without invoking ``the first reason" given above. Note again, that all of the preceding (including discussion of the smallness of the Higgs portal mass contribution) is equally applicable to the quartic curvaton, which requires a quartic self-coupling about an order of magnitude smaller than the inflaton.

\bibliographystyle{JHEP.bst}

\bibliography{lightinflaton}

\end{document}